\documentclass[twocolumn,aps,pra,superscriptaddress, 10pt]{revtex4-2}
\usepackage[utf8]{inputenc}
\usepackage[T1]{fontenc}
\usepackage[english]{babel}
\usepackage{bm}
\usepackage[dvipsnames]{xcolor}
\usepackage{amsfonts}
\usepackage{ulem} 
\usepackage{tcolorbox}
\usepackage{wrapfig} 
\usepackage{hyperref}
\usepackage{soul}
\usepackage{physics}
\AtBeginDocument{\RenewCommandCopy\qty\SI}
\usepackage{lineno}
\usepackage{tabularray}
\usepackage[ruled,lined]{algorithm2e}
\usepackage{subcaption}

\usepackage{lipsum}

\hypersetup{
colorlinks=true,
citecolor=blue,
linkcolor=red,
urlcolor=blue
 ,
 pdfmenubar=true
}

\usepackage{caption}
\newcommand{\bfx}{\mathbf{x}}

\setstcolor{blue}

\begin{document}

\title{Learning the dynamics of Markovian open quantum systems from experimental data}

\author{Stewart Wallace}
\affiliation{ 
SUPA, Institute of Photonics and Quantum Sciences, School of Engineering and Physical Sciences, Heriot-Watt University,
Edinburgh EH14 4AS, UK
}

\author{Yoann Altmann}
\email{y.altmann@hw.ac.uk}
\affiliation{ 
Institute of Signals, Sensors and Systems, School of Engineering and Physical Sciences, Heriot-Watt University,
Edinburgh EH14 4AS, UK
}

\author{Brian D. Gerardot}
\affiliation{ 
SUPA, Institute of Photonics and Quantum Sciences, School of Engineering and Physical Sciences, Heriot-Watt University,
Edinburgh EH14 4AS, UK
}

\author{Erik M. Gauger}
\email{e.gauger@hw.ac.uk}
\affiliation{ 
SUPA, Institute of Photonics and Quantum Sciences, School of Engineering and Physical Sciences, Heriot-Watt University,
Edinburgh EH14 4AS, UK
}

\author{Cristian Bonato}
\email{c.bonato@hw.ac.uk}
\affiliation{ 
SUPA, Institute of Photonics and Quantum Sciences, School of Engineering and Physical Sciences, Heriot-Watt University,
Edinburgh EH14 4AS, UK
}

\begin{abstract}
We present a Bayesian algorithm to identify explainable processes generating open quantum system dynamics, described by a Lindblad master equation, that are compatible with measured experimental data. The algorithm, based on a Markov Chain Monte Carlo approach, assumes the energy levels of the system are known and outputs a ranked list of interpretable master equation models that produce predicted measurement traces that closely match experimental data. We benchmark our algorithm on quantum optics experiments performed on single and pairs of quantum emitters. The latter case opens the possibility of cooperative emission effects and additional complexity due to the possible interplay between photon and phonon influences on the dynamics. Our algorithm retrieves various minimal models that are consistent with the experimental data, and which can provide a closer fit to measured data than previously suggested and physically expected approximate models. Our results represent an important step towards automated systems characterization with an approach that is capable of working with diverse and tomographically incomplete input data.  This may help with the development of theoretical models for unknown quantum systems as well as providing scientists with alternative interpretations of the data that they might not have originally envisioned and enabling them to challenge their original hypotheses.  
\end{abstract}

\maketitle

\section{\label{sec:intro} Introduction}

Automating the extraction of physical models from experimental data in a reliable, robust, and interpretable form is an important challenge with far-reaching implications for both fundamental scientific discoveries and technological applications \cite{butler_machine_2018, 
 carleo_machine_2019, moosavi_role_2020, 
 krenn_scientific_2022, gebhart_learning_2023, krenn_artificial_2023}. In fundamental science, artificial intelligence can be used by scientists to explore quantitative models, through inference on experimental data, and study their interpretation. In particular, it may be used to propose viable alternative interpretations that researchers may not have initially envisioned as being relevant when designing the experiments. Automated model identification and selection could also be useful when characterizing multiple versions of a system, for example with changes in composition (either randomly occurring or intentionally introduced), or subject to variations in external conditions \cite{usman_framework_2020, moon_machine_2020, lee_multiple_2021, valenti_scalable_2022, thomas_rapid_2023, craig_bridging_2024}. In the future, this might lead to an automated research cycle, with machines developing models for physical systems, and designing and executing optimal experiments to discriminate between competing theories. 

Stimulated by these motivations, researchers have designed procedures to automate the construction of models for classical \cite{schmidt_distilling_2009, liu_machine_2021, lemos_rediscovering_2023, chen_constructing_2024} and quantum systems \cite{torlai_learning_2016, van_nieuwenburg_learning_2017, carrasquilla_reconstructing_2019, torlai_integrating_2019, gentile_learning_2021, anshu_sample-efficient_2021, flam-shepherd_learning_2022, arlt_digital_2022, huang_provably_2022, craig_bridging_2024}. 
A particular challenge for the construction of quantum models is that most quantum systems interact substantially with their wider physical environment and cannot be well approximated as closed systems. Some such interactions are crucial for controlling and measuring quantum systems, whilst others are undesirable yet always present. This necessitates operating in the paradigm of open quantum system dynamics where the influence of the wider physical environment is captured through additional incoherent processes \cite{breuer_theory_2007}. 

A general way of learning how an open quantum system evolves between two fixed points in time is quantum process tomography \cite{b_p_lanyon_efficient_2017, granade_practical_2016}. An extension to multi-time settings is possible in the framework of process tensors \cite{pollock_non-markovian_2018}, albeit tomographic reconstructing of those comes with daunting resource requirements \cite{White_PTT_22}, and it can be challenging to extract insight and information about dynamical (noise) processes \cite{Pollock_PTMarkov,cygorek2024understanding}. Beyond resource cost, tomographic methods are limited to (few) discrete time steps and require extensive pre-existing knowledge about the system, as one needs to be able to prepare a set of fiducial initial states and to perform a complete set of quantum gates with high-fidelity. They are therefore most useful for characterizing well-controlled quantum processes and circuits, such as small-scale quantum processors \cite{samach_lindblad_2022} or quantum sensors, but are less suited to exposing the dynamics of largely unknown quantum systems under limited control. 

\begin{figure*}[!htbp]
\centering
\includegraphics[width=1\textwidth]{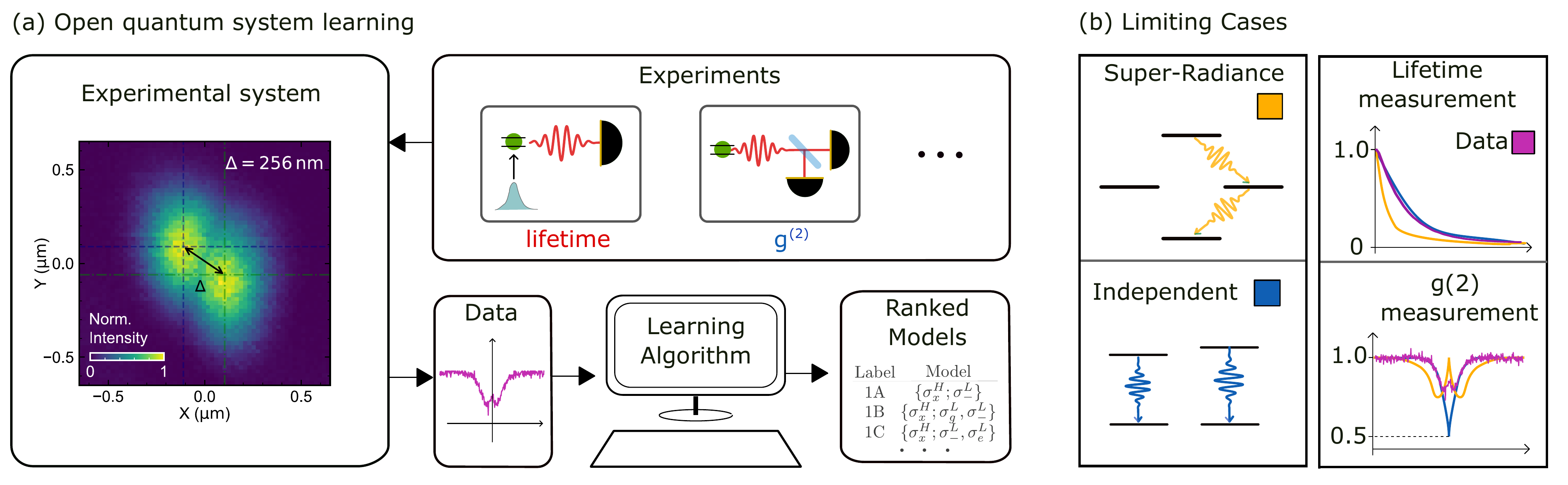}
\caption{\textbf{a) Overview of the procedure to learn the dynamics of an unknown open quantum system.} An experimental system is probed by a set of experiments involving optical excitation and the detection of the emitted photons, with the goal to determine the processes governing the internal dynamics, and the related rates.  The learning algorithm autonomously builds combinations of Hamiltonian and Lindbladian operators, ranked based on their ability to fit the data through a master equation model. We apply our algorithm to the case of two self-assembled semiconductor quantum dots separated by $\delta  =256$nm, optically addressed simultaneously. We consider experimental data by Koong et al \cite{koong_coherence_2022}, where the system is interrogated with two types of experiments: lifetime measurements, which probe the decay of an optically-excited state, and auto-correlation $g^{(2)}$ measurements, which detect correlations between the detected photons' arrival times. 
\textbf{b) Idealized limiting cases.} For two ideal indistinguishable emitters in close proximity, one could expect super-radiant decay, evidenced (orange curves) by a change of the emission lifetime into a bi-exponential decay, with simultaneous emergence of an anti-dip near $g^{(2)} (\tau) \sim 0$. As another limit case, two independent emitters (blue curves) would not show any change in lifetime, with a $g^{(2)} (\tau)$ dip down to $g^{(2)} (\tau = 0) \geq 0.5$. The experimental data (purple curves) are not described by either of the two ideal limiting cases.
}
\label{fig:experimental figure}
\end{figure*}

An attractive alternative to tomographic reconstruction of quantum channels would be to learn the dynamical generator of the system of interest \cite{breuer_theory_2007}. Generally, this may also feature a high level of complexity and be non-Markovian, i.e.~depend on the history of the system and environment, but an important subclass are time-local and can be captured by a Gorini–Kossakowski–Sudarshan–Lindblad (GKSL, or for brevity simply Lindblad) generator \cite{vittorio_gorini_completely_1976,lindblad_generators_1976}.
For such open quantum systems operating in the Markovian regime, quantum channel tomography can, for example, be replaced by pertinent terms of a Lindblad master equation \cite{samach_lindblad_2022}.

A paradigmatic example of Markovian open quantum system dynamics is the interaction between light and matter, a topic of fundamental scientific and technological importance. Specifically, light-matter interaction encompasses phenomena that can be simple enough to be entirely describable by understandable Lindblad processes, yet complicated enough to display rich physics and present interesting quantum effects \cite{gammelmark_bayesian_2013}, especially if multiple emitters are present and cooperative effects emerge. Experimentally, single and few emitter systems can nowadays readily be probed via the standard quantum optics toolkit, for instance through the detection of photon arrival times, spectroscopy, and via correlations of emitted photons \cite{marc_asmann_measuring_2010, kim_super-radiant_2018,d_levonian_optical_2021, koong_coherence_2022, rebecca_e_k_fishman_photon-emission-correlation_2023, cygorek_signatures_2023, daniil_m_lukin_two-emitter_2023, cygorek_signatures_2023, tiranov_collective_2023}. For these reasons, we shall presently focus on this platform to test our open quantum system learning algorithm. 

Recently, researchers have been investigating how data-driven and deep learning techniques can be used to model quantum systems. Deep learning approaches \cite{banchi_modelling_2018, hartmann_neural-network_2019, luchnikov_machine_2020, mazza_machine_2021} are appealing as they can provide compact expressive parametrizations for quantum states whose exact analytical (and sometimes even numerical) description may be unfeasible due to the large number of degrees of freedom necessary to describe them. In the context of open quantum systems, neural networks have been used to model dynamics. Examples are the use of recurrent neural networks to learn the dynamics of non-Markovian systems \cite{banchi_modelling_2018, krastanov_unboxing_2020} and local generators, such as the Liouville operator \cite{carnazza_inferring_2022, cemin_inferring_2024} and extrapolate dynamics to later (unseen) times \cite{mazza_machine_2021}. In these works, the approximation of the generators of dynamics by a neural network provides a degree of interpretability, in contrast for example to quantum process tomography. One of the main limitations of most deep learning approaches is that neural networks are often used as ``black boxes'' with limited connection to explainable physical processes. Although progress is being made towards creating more interpretable deep learning approaches, such as learning symbolic physics with graph networks \cite{miles_cranmer_discovering_2020} or through operationally meaningful representations \cite{nautrup_operationally_2022}, understanding physics encoded in neural networks remains a challenge.

In this paper, we demonstrate an algorithm to address the problem of automatically learning descriptions of open quantum systems within a Lindbladian framework, from a set of experimental data, with minimal prior knowledge and assumptions. We do not assume the availability of a complete set of fiducial initial states and high-fidelity gates, making the algorithm suitable to exploring the dynamics of unknown quantum systems through a  tomographically-incomplete set of experiments. We leverage the power of Bayesian inference \cite{gammelmark_bayesian_2013}, which facilitates explainability of the retrieved models and enables assessing uncertainties of the results~\cite{Craig2024}. Previous work has addressed the learning of open quantum systems dynamics by retrieving the (Liouville) generators of the dynamics \cite{carnazza_inferring_2022, cemin_inferring_2024}, an important step towards improved interpretability of learned models. Here, we go even further towards explainability, by targeting representations in terms of intuitively understandable elementary physical processes and developing an algorithm that learns elementary Lindblad processes with direct physical meaning.

We initially benchmark our algorithm on the simple case of a single emitter under coherent excitation, and then apply it to characterize the optical emission dynamics of two resonant quantum emitters. By setting a penalty on higher model complexity, the algorithm succeeds in finding minimal Lindblad models for the two-emitters system. These postulate an asymmetry in the optical excitation/decay pathways and reveal that dephasing is not necessarily required to explain the observed physics. Although linking the retrieved models to novel explainable physics is not straightforward, our algorithm can play a role in inspiring researchers to consider alternative explanations for data from experiments designed with one specific model in mind.    

\begin{figure*}[!htbp]
\centering
\includegraphics[width=1\textwidth]{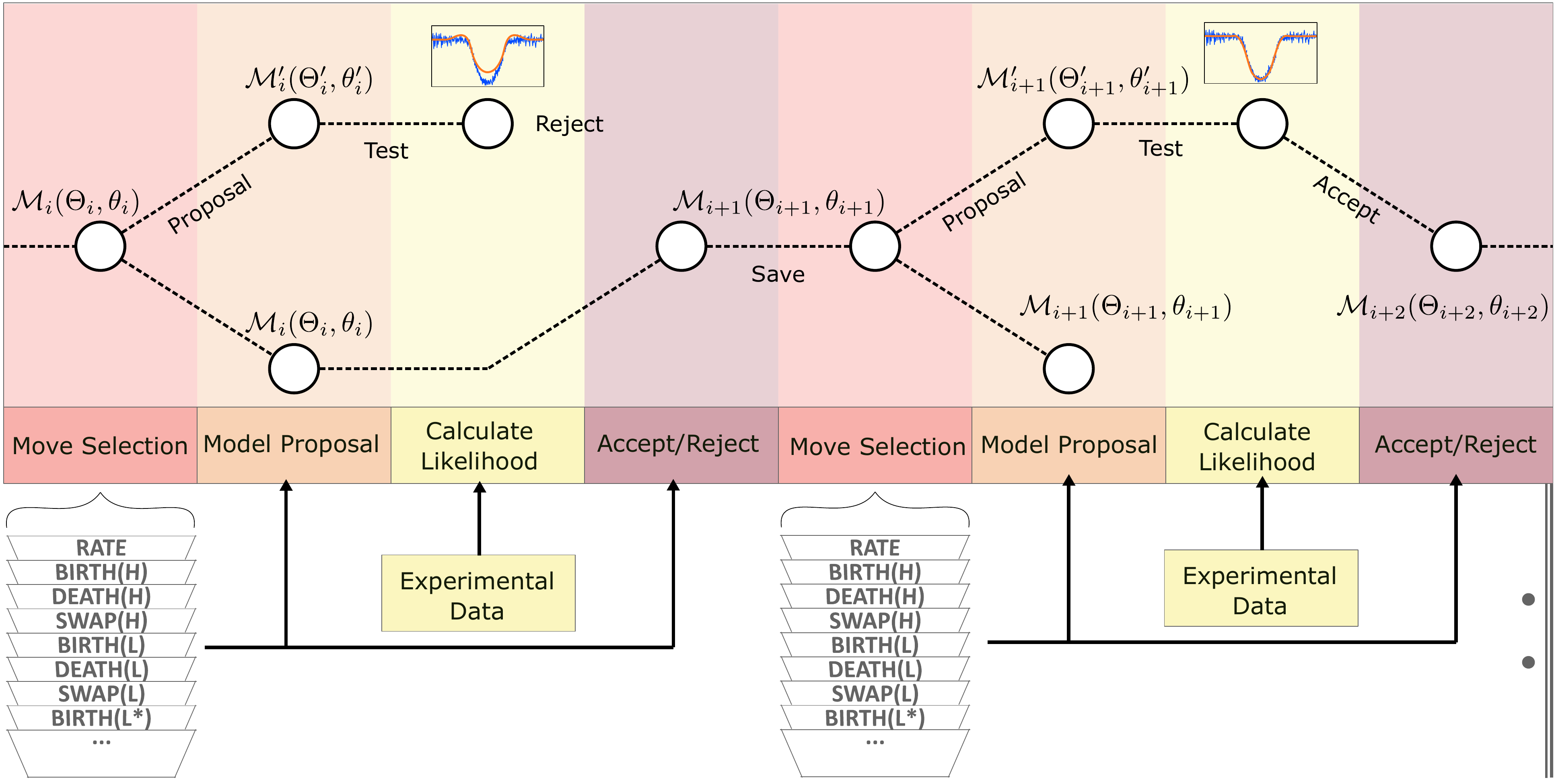}
\caption{ \textbf{Sketch of the learning algortihm.}
The learning algorithm performs a sequential exploration of models ($\mathcal{M}_i(\boldsymbol{\Theta}_i,\boldsymbol{\theta}_i)$), described by a combination of Hamiltonian and Lindbladian operators $\boldsymbol{\Theta}_i$, in a master equation (Eq.~\ref{eq:lindblad}), and the associated rates $\boldsymbol{\theta}_i$. At each step of the algorithm, a new model $\mathcal{M}_{i+1}(\boldsymbol{\Theta}_{i+1},\boldsymbol{\theta}_{i+1})$ is proposed, by applying one of the allowed moves (e.g.~parameter optimization, add Hamiltonian or Lindbladian term, remove Hamiltonian or Lindbladian term, etc), chosen at random from a library. The new model is then compared to the existing model $\mathcal{M}_i(\boldsymbol{\Theta}_i,\boldsymbol{\theta}_i)$, in terms of their performance in explaining the experimental data, and accepted or rejected. This procedure creates Markov chains of models, which are then converted to the Liouvillian formalism and clustered into model classes $\mathcal{\hat{M}}_q$, ranked according to how often the models in each class appear in the chains (as sampling frequency is proportional to likelihood).
A detailed description of the algorithm can be found in  Appendix \ref{app:algorithm_overview}.}
\label{fig:algorithmic figure}
\end{figure*}

\section{Results}
\label{Sec: Results}
\subsection{Approach}
Our goal is the construction of models described by the GKSL formalism~\cite{vittorio_gorini_completely_1976,lindblad_generators_1976} from experimental data. The GKSL approach, used widely in the field of open quantum dynamics, guarantees physicality of the resulting dynamical map, and captures the Markovian dynamics of the reduced density matrix of a quantum system that is embedded in a memory-less environment, such as a structureless electromagnetic environment. Its canonical form reads ($\hbar=1$)  
\begin{equation}
	 \dot{\rho} = -i[H,\rho] + \sum_j^{n_l} \gamma_j \left(L_j \rho L_j^\dagger - \{ L_j^\dagger L_j,\rho\}/2\right),
	\label{eq:lindblad} 
\end{equation} 
where $H$ is the system Hamiltonian, $L_j$ are $n_l$ Lindblad operators that describe dissipative processes at rates $\gamma_j$, and $\{A,B\}=AB+BA$ denotes the anti-commutator.
A challenge for the learning of such a model is that we do not know a priori how many distinct Lindblad operators $n_l$ are required. Each such operator $L_j$ is a $d\times d$ matrix,  where $d$ is the dimension of the system Hilbert space, and the non-linearity of Eq.~\eqref{eq:lindblad} entails a large number of possible non-equivalent combinations.
Eq.~\eqref{eq:lindblad} can be mapped to Liouville space as $\frac{d}{dt}\vert \rho \rangle = {\mathcal{L}} \vert \rho \rangle$, where $\vert \rho \rangle$ is a vectorized density matrix and where the Liouvillian operator ${\mathcal{L}}$ has $d^2 \times d^2$ complex elements. This has the formal solution $\vert \rho(t) \rangle = e^{{\mathcal{L}} t} \vert \rho(0)\rangle$, and learning $\mathcal{L}$ would thus be an alternative to a reconstruction of Eq.~\eqref{eq:lindblad}. Learning a Lindblad master equation in the GKSL formalism can, however, be preferable for explainability, as it provides a direct physical interpretation of the most relevant noise processes (e.g. photon emission processes). Further, an important feature of the GSKL formalism is that it defines a dynamical semigroup ensuring completely positive and trace preserving dynamics, whereas a general Liouvillian provides no such guarantee.  Note that, while the mapping from Eq.~\eqref{eq:lindblad} to $\mathcal{L}$ is straightforward, the reverse is neither unambiguous nor necessarily possible. 

The overarching idea of this work is presented in Fig. ~\ref{fig:experimental figure}. Given an unknown quantum system, we perform a set of controlled, but possibly tomographically incomplete, experiments. The learning algorithm processes the resulting experimental data, without any real-time feedback, and finds a ranked list of GKSL equations that are compatible with the data. While the algorithm we propose below is in principle applicable to general Markovian systems, we consider here the specific case of the optical emission from one or both of two proximal ($256.1$ nm separation) self-assembled InGaAs quantum dots (QDs), see Fig. ~\ref{fig:experimental figure} a), as reported by Koong et al ~\cite{koong_coherence_2022}. Each of the individual quantum dots can be modelled as a two-level system (2LS), whose energy levels can be tuned by an applied electric field, through the DC Stark effect. This enables experimental control of the transition between a situation where the dots' optical emissions are frequency-detuned from each other, behaving as completely independent emitters, and a situation where they are resonant with each other, potentially leading to effects like super-radiance or cooperative emission (Fig.~\ref{fig:experimental figure} (b)) \cite{cygorek_signatures_2023,wiercinski_phonon_2023}. 

We consider experimental data from two types of experiments on the same physical system:
lifetime measurements (corresponding to dataset $\tilde{y}_{LT} (\tau)$) and second-order intensity correlations 
(dataset $\tilde{y}_{g^{(2)}} (\tau)$), both dependent on the arrival time $\tau$ of photons under optical excitation. In lifetime measurements, the emitters are excited with a short laser pulse, and the arrival time $\tau$ of the emitted photons is detected and plotted in a histogram that tracks the temporal decay of the excited state. Independent emitters maintain the same exponential decay timescale 
of a single emitter, while super-radiant emitters exhibit an emitted intensity profile that departs from mono-exponential behaviour~\cite{tiranov_collective_2023}.

In intensity correlation ($g^{(2)}$) measurements, emitted photons are sent to a beamsplitter and redirected to two single-photon detectors: the measured signal consists of coincidence counts between the two photo-detectors, as a function of the time delay between  photon detections. For two independent emitters, one expects a zero-time-delay dip down to $1/2$, i.e.~$g^{(2)}(0) = 0.5$, with the recovery towards the long delay baseline of $g^{(2)}(\abs{\tau}\to\infty) =1$ determined by the lifetime, decoherence, and driving conditions. By contrast, 
in the super-radiant regime a two term exponential lifetime decay would be expected with an associated anti-dip in the $g^{(2)}$ at short time delays that can reach, or exceed unity at $\tau = 0$~\cite{auffeves_few_2011, cygorek_signatures_2023}. It is clear that neither of these two idealized limiting cases---independent or super-radiant emission---can explain the experimental data in Fig.~\ref{fig:experimental figure}(b).

To develop our algorithm, we assume the Hilbert space dimension and the emitter level structure is known. This especially includes knowledge about allowed optical transitions, for example from spectroscopic measurements. We will here consider a single and a double emitter system with Hilbert space dimension $d=2$ and $d=4$, respectively. The goal then is to find a model $\mathcal{M}(\boldsymbol{\Theta},\boldsymbol{\theta})$, defined as a combination of Hamiltonian and Lindbladian processes $\boldsymbol{\Theta} = \lbrace {H}_h, h=1..n_h; {L}_l, l = 1... n_l\rbrace$ and their associated energies and rates $\boldsymbol{\theta}=\lbrace \omega_h, h=1...n_h; \gamma_l, l = 1...n_l \rbrace$, which explains the experimental data. Note that as $\mathcal{M} (\boldsymbol{\Theta},\boldsymbol{\theta})$ is fully characterized by $\boldsymbol{\Theta}$ and $\boldsymbol{\theta}$, we simply use the notation $\mathcal{M}$ in the remainder of the paper unless the explicit notation is needed.

We now briefly discuss the operator space of our system. As explained above, the number of possible Lindblad operators is large, even for our relatively small system. 
To facilitate interpretability of the learned dynamics and keep the model space tractable, we restrict the space of potential operators as explained below. We consider the canonical basis of $2 \times 2$ matrices for a single 2LS,
\begin{align}
\label{eq:library}
    {\sigma}_+ = \begin{bmatrix}
        0 & 1 \\
        0 & 0 \\
    \end{bmatrix} \,,
    \hspace{3pt} 
    {\sigma}_- = \begin{bmatrix}
        0 & 0 \\
        1 & 0 \\
    \end{bmatrix} \,,
    \hspace{3pt}
    {\sigma}_e = \begin{bmatrix}
        1 & 0 \\
        0 & 0 \\
    \end{bmatrix} \,,
    \hspace{3pt}
    {\sigma}_g = \begin{bmatrix}
        0 & 0 \\
        0 & 1 \\
    \end{bmatrix} \,,
\end{align}
alongside their pairwise tensor products $\sigma_{ij} = \sigma_i \otimes \sigma_j$, where $i,j \in \{+, -,g, e\}$, to arrive at 16 unique $4 \times 4$ operators for the two emitter problem. So far, our library is just the canonical basis set for the full Hilbert space of our system, in both cases. We now limit the number of possible operators that can be constructed from this `library' by only allowing equally weighted linear combinations of up to $\mathcal{C}$ basis operators. For $d=4$ and $\mathcal{C}=2$ this gives $\sum_{c=1}^{\mathcal{C}} \binom{d^2}{c}=16+120$ such combinations. The choice $\mathcal{C}=2$ is motivated by allowing for a sufficiently rich set of operators, without leading to a number that becomes overly computationally unwieldy. Once one removes degenerate combinations arising from terms such as $\sigma_g+\sigma_e = \sigma_e+\sigma_g$, we are left with a manageable number of 105 unique processes. 
By contrast, for $d=2$ and $\mathcal{C}=2$ we only have $\binom{4}{1} + \binom{4}{2} = 10$  combinations, and in this special case this includes the identity operator. More details can be found in Appendix \ref{app:operator_space}.

As the number of processes involved in the system is typically unknown a-priori, we cast the process identification task as a model selection/comparison problem. We develop an algorithm based on reversible-jump Markov-chain Monte Carlo (rj-MCMC) \cite{peter_j_green_reversible_1995}, i.e., an extension of more traditional Markov Chain Monte Carlo (MCMC) methods. rj-MCMC is specialized at exploring spaces of varying dimensionality, with or without structural overhaul. Thus it is ideal for identifying processes that are present in a model, and the rate at which these processes act.

The algorithm starts from an initial model consisting of a random set of processes, both Hamiltonian and Lindbladian, and random associated rates. As sketched in Fig.~\ref{fig:algorithmic figure}, we then create a chain of models, where at each iteration $i$, a new candidate model is proposed randomly through a set of moves, e.g. adding a process (either Hamiltonian or Lindbladian), removing a process, tuning the rate of a given process, etc. All moves are described in details in Appendix \ref{app:algorithm_overview}. The plausibility of the new candidate $\mathcal{M}_i$ is compared to that of the most recently accepted candidate $\mathcal{M}_{i-1}$ (see Appendix) and the candidate model is accepted or rejected according to the rj-MCMC standard Metropolis-Hastings-like acceptance rate \cite{peter_j_green_reversible_1995} (Eq. ~\ref{Eq: accept reject}). 

To better explore the model space and confirm convergence, we run a set of parallel chains, each from random initial conditions, for a pre-determined number of steps. The models generated within all chains are converted to Liouvillian matrices (as these represent a unique description of a given physical model). The Liouvillians are clustered into model classes ($\mathcal{\hat{M}}_{q}$), each representing different GKSL master equation models corresponding to similar physics (see Appendix \ref{app:ranked_list}). We analyze the model classes in terms of their ``popularity'', corresponding to how often the related models are present in the chains (as sampling frequency is proportional to likelihood).

\begin{figure*}[!htbp]
\includegraphics[width=0.95\textwidth]{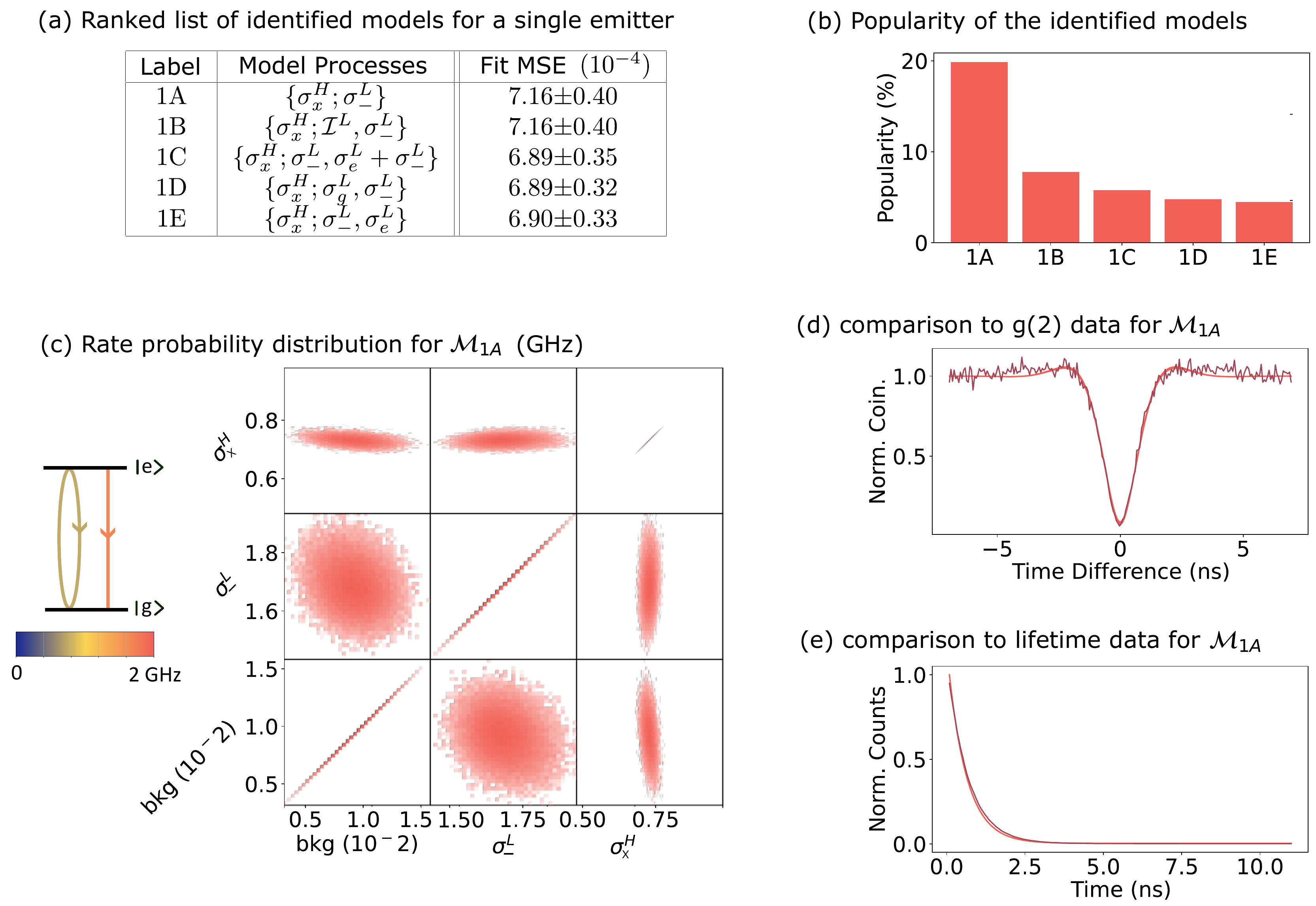}
\title{Single Emitter Learning}
\caption{\textbf{Model learning for a single emitter under resonant coherent optical excitation.} \textbf{a)} Top five models found by our algorithm, ranked in terms of popularity, listed with the corresponding mean squared error (MSE) of the fit. For each model, we list the Hamiltonian (super-script ``H'') and Lindbladian terms (super-script ``L'') involved. All five models are found to be part of the same model class $\mathcal{\hat{M}}_{1}$ (see main text). \textbf{b)} Histogram of the popularity of the top five models in $\mathcal{\hat{M}}_{1}$. \textbf{c)} On the left, sketch of the processes for $\mathcal{M}_{1A}$ (circle for Hamiltonian processes, straight line for Lindbladian ones), with the colour indicating the relative rate, as described by the adjacent colourbar. On the right, set of scatter plots showing 2D projections of the joint probability distribution for the three rates associated to $\mathcal{M}_{1A}$: the two rates associated with the two processes $\sigma_+^H$ and $\sigma_-^L$, and the experimental background rate. A single cluster is evident in the distribution, showing there are no degenerate rates within this model class. The distribution is sampled with MCMC using $\sim 1.2 \times 10^{6}$ samples. \textbf{d)} Comparison between experimental (gray curve) $g^{(2)} (\tau)$ and simulations (red) based on model $\mathcal{M}_{1A}$. Simulations are based on 5000 samplings of the distribution: we plot the mean, with a shaded region (hardly visible) corresponding to one standard deviation. \textbf{e} Comparison between experimental data (gray) and simulations (red) for lifetime measurements, plotted as in (d). 
}
\label{fig:single_emitter}
\end{figure*}

\subsection {Outcomes: single quantum emitter}
To illustrate the performance of our approach, we first benchmark the algorithm to reconstruct models for photon emission dynamics for a single two-level emitter under resonant optical excitation (Fig.~\ref{fig:single_emitter}). The ground and excited states of the system are respectively labeled as $\ket{g}$ and $\ket{e}$. For this experiment we use 60 chains, each with 10$^5$ models, prior to burn-in post selection.

We list in Fig.~\ref{fig:single_emitter}(a) the five most popular models found by our algorithm, all belonging to one single model class $\mathcal{\hat{M}}_1$.
For each model 1A-1E, we list the associated Hamiltonian and Lindbladian processes, and the mean square error (MSE) between the experimental data and the data simulated from each model, as a mean figure of merit for the quality of the fits. As all models belong to one single cluster, they feature similar values for the MSE. Note that the MSE only measures the quality of the fit, without taking the complexity of the model into account. The relative popularity for each model within each cluster is presented in the bar chart in Fig. ~\ref{fig:single_emitter}(b). The relative popularity is a used here as a proxi for the Bayes factors to compare models, and is computed  by averaging the time spent exploring each model during the Markov chains. It is interesting to note that the most popular models (1A and 1B) present marginally higher MSEs than the subsequent models. This illustrates the capability of our method to prevent overfitting (preferring simpler models).

Model $\mathcal{M}_{1A}$ contains the terms one would intuitively expect for a single two-level system under resonant excitation, i.e.~a Hamiltonian Pauli operator ${\sigma}^H_x$, which describes coherent driving, and one Lindbladian incoherent decay term ${\sigma}^L_-$. $\mathcal{M}_{1B}$ is identical to $\mathcal{M}_{1A}$ with the trivial addition of an inconsequential identity Lindbladian~\footnote{The identity is possible for a single 2LS only, when allowing operator complexity up to a fixed $\mathcal{C}$ (in this work $\mathcal{C}=2$). Its appearance increases confidence in $\mathcal{M}_{1A}$ being the top model while also demonstrating our approach does not penalize additional processes to the extent that might suppress their inclusion if relevant.}.
Model $\mathcal{M}_{1C}$ has an additional term $\sigma_e^L + \sigma_-^L$, which incoherently transitions population from the $\ket{1}$ into the superposition state $\ket{0} + \ket{1}$. This transition does not register as a photon detector click, due to our heuristic (Appendix \ref{app: Emission and Excitation Detection}) on optically allowed transitions. This model has a slightly lower MSE than $\mathcal{M}_{1A}$-$\mathcal{M}_{1B}$: it fits the experimental data slightly better but is penalized in popularity due to the model prior. A similar effect is seen for models $\mathcal{M}_{1D}$ and $\mathcal{M}_{1E}$, which add dephasing operators. 

Model $\mathcal{M}_{1A}$ includes a total of three rates: the two rates associated with the learned processes (Hamiltonian driving $\sigma_x^H$ and Lindbladian decay $\sigma_-^L$), and the background count rate. As a 3-dimensional probability distribution is difficult to visualize in two dimensions, we show the projected distributions for each pairs of rates in Fig.~\ref{fig:single_emitter}(c). The probability distribution for the rates is computed through parameter learning (i.e. running a standard MCMC on the specific set of processes learned by our algorithm) and presents one single cluster, confirming that the model has a single set of likely rates, as one would expect from a simple coherently pumped system. 
A comparison between the predictions of model $\mathcal{M}_{1A}$ (red curve) and experimental data (gray curve) is shown in Fig.~\ref{fig:single_emitter}(d) ($g^{(2)}(\tau)$ measurement) and Fig.~\ref{fig:single_emitter}(e) (lifetime measurement), with the hardly noticeable shaded area representing the standard deviation.

\begin{figure*}[!htbp]
\centering
\includegraphics[width=0.95\textwidth]{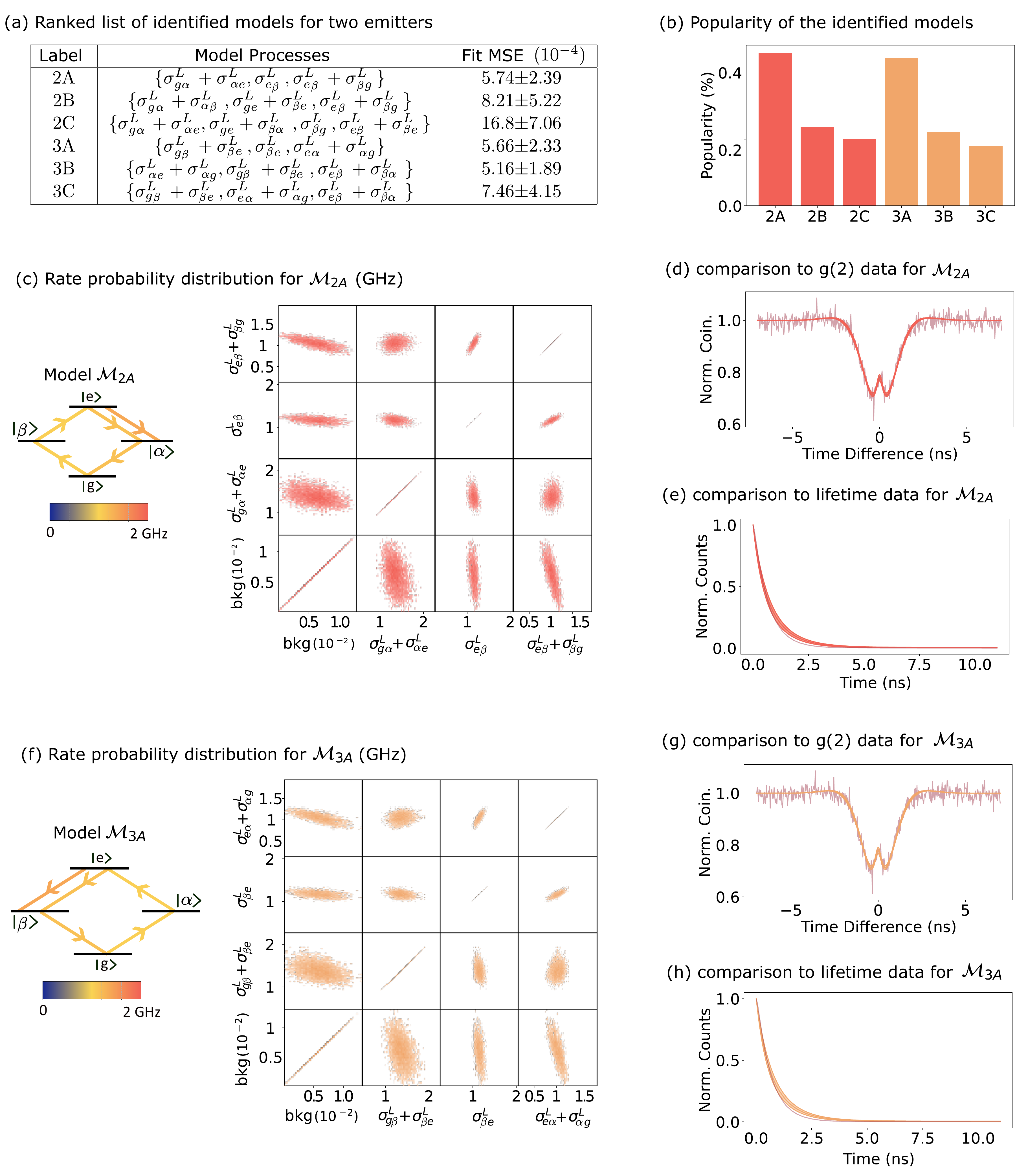}
\caption{ \textbf{Model learning for two emitters under incoherent excitation.} \textbf{a)} The algorithm finds two model classes  $\mathcal{\hat{M}}_2$ and $\mathcal{\hat{M}}_3$, each including $>400 \times 10^3$ models. The models are ranked in terms of popularity (shown in the histogram in \textbf{(b)}): here we list the top three models from each class, with corresponding Hamiltonian and Lindbladian operators and alongside the mean squared error (MSE) of the fit. Sub-figures \textbf{c)} and \textbf{f)} show scatter plots representing 2D projections of the joint probability distribution for the rates associated respectively to $\mathcal{M}_{2A}$ ( $\lbrace \sigma_{g \alpha}^L + \sigma_{\alpha e}^L, \sigma_{e \beta}^L + \sigma_{\beta g}^L, \sigma_{\alpha e}^L \rbrace$, background count rate) and $\mathcal{M}_{3A}$ ( $\lbrace \sigma_{g \alpha}^L + \sigma_{\alpha e}^L, \sigma_{e \beta}^L + \sigma_{\beta g}^L, \sigma_{\alpha e}^L  \rbrace$, background count rate). Sub-figures \textbf{c)} and \textbf{f)} also include, on the left, a sketch of the involved processes (all Lindbladian, straight arrows) with rates encoded in the the colour of the arrow. Sub-figure \textbf{d)} (\textbf{g)}) shows a comparison between experimental (gray curve) $g^{(2)} (\tau)$  and simulations (red curve) based on model $\mathcal{M}_{2A}$ ($\mathcal{M}_{3A}$), while \textbf{e)} (\textbf{h)}) displays the same comparison for the lifetime measurements. Simulations in \textbf{c)}, \textbf{d)}, \textbf{g)}, \textbf{h)} are based on 5000 samplings of the distribution: we plot the mean, with a shaded region (hardly visible) corresponding to one standard deviation. 
}
\label{fig:double_emitter}
\end{figure*}

\subsection {Outcomes: two quantum emitters}

We next apply our algorithm to the case of two two-level emitters which we shall treat as a generic `diamond'-shaped four-level system. We label its four states as $\ket{g}$ (in terms of single particle states this would be $\ket{gg} \equiv \ket{g}\otimes \ket{g}$) for the ground and $\ket{e}$ (short for $\ket{ee}$) for the doubly excited state. Further, we assign the labels $\ket{\alpha}$ and $\ket{\beta}$ to the two singly excited levels. We deliberately do not impose a specific relationship between $\ket{\alpha}$ and $\ket{\beta}$ and the canonical basis; $\ket{\alpha}$ and $\ket{\beta}$ could thus, for instance, just be the basis states $\ket{eg}$ and $\ket{ge}$, or the symmetric and antisymmetric states (i.e.~$(\ket{eg} + \ket{ge})/\sqrt{2}$ and $(\ket{eg} - \ket{ge})/\sqrt{2}$), or indeed any other (orthogonal) set of states in this subspace. This allows our learning algorithm to automatically adopt the most appropriate basis that features the most compact representation. Optical transition rates that link the intermediate levels $\ket{\alpha}$ and $\ket{\beta}$ to the ground and doubly excited states, respectively, are available via the library of $105$ unique operators discussed above. However, we emphasize that we now build on the operator \textit{matrices} (rather than making reference to their original construction as a tensor product of canonical basis operators). In other words, the operator matrix representation now acts on the flexibly adopted basis of $\{ \ket{g}, \ket{\alpha}, \ket{\beta}, \ket{e}\}$ (for which the relationship with $\{\ket{g}, \ket{ge}, \ket{eg}, \ket{e}\}$ remains unspecified).
As this two-emitter problem is significantly more complex than our previous single emitter example, we extend the number of steps in the Markov chains by a factor 10, to $10^6$ steps. This yields the results presented in Fig.~\ref{fig:double_emitter}. 

The top two model classes identified by our algorithm, labeled $\mathcal{\hat{M}}_{2}$ and $\mathcal{\hat{M}}_{3}$, are listed in Fig.~\ref{fig:double_emitter}(d). They both feature an excellent fit to the experimental data (Fig.~\ref{fig:double_emitter}(d)/(e) and Fig.~\ref{fig:double_emitter}(g)/(h), respectively). As in the schematics in Fig.~\ref{fig:double_emitter}(c) and (f), both model classes represent essentially identical physics, becoming largely equivalent upon relabeling the two intermediate singly-excited states: $\mathcal{\hat{M}}_{2}$ features the terms ${\sigma}^L_{e\alpha}+{\sigma}_{\alpha g}^L$ and ${\sigma}^L_{g \beta}+{\sigma}^L_{\beta e}$ whereas ${\sigma}^L_{e \beta}+{\sigma}_{\beta g}^L$ and ${\sigma}^L_{g \alpha}+{\sigma}^L_{\alpha e}$ are exclusively present in $\mathcal{\hat{M}}_{3}$. Clearly, these terms are identical upon swapping the arbitrarily assigned state labels `$\alpha$' and `$\beta$'; and this symmetry is also reflected in the nearly identical-looking joint probability distributions shown in Fig.~\ref{fig:double_emitter}c and \ref{fig:double_emitter}f. Further, there are shared terms present across both classes: models $\mathcal{M}_{2B},\mathcal{M}_{2C},\mathcal{M}_{3B},\mathcal{M}_{3C}$ each contain a process that relaxes the system without the registered emission of a photon, $\sigma^L_{ge}+\sigma^L_{\alpha e}$, $\sigma^L_{ge}+\sigma^L_{\alpha g}$, $\sigma^L_{e \alpha}+\sigma^L_{\alpha \beta}$, $\sigma^L_{e \alpha}+\sigma^L_{\alpha \beta}$, respectively. Such relaxation terms are not dipole-allowed, and we therefore consider them as dark relaxation that does not register as an emission event in $g^{(2)}$ or lifetime simulations. Further, some models include (also optically dark) transitions between the states in the single excitation manifold through $\sigma^L_{\alpha \beta}$ or $\sigma^L_{\beta \alpha}$ terms, while others instead feature dephasing between different excitation manifolds. 

A notable feature of our models is the preference to separate the excitation from the emission branch, combined with a relatively faster emission from the doubly-excited to a singly-excited manifold alongside an emission term that spans both optical transitions. Neither of the two top model classes obviously maps onto the archetypal two-emitter scenarios depicted in Fig.~\ref{fig:experimental figure}(b). Specifically, neither model class supports a description based on superradiance (as expected by the fact that the emitters lifetime does not change compared to the single-emitter case), nor are they representative of entirely independent emission (which cannot explain the anti-dip seen in the $g2$ experiments). While our learned models consist of simple Lindblad terms and provide very good match to the data, their interpretation in terms of physical processes underpinning the observed quantum emitter dynamics is not straightforward.

\begin{figure*}[!htbp]
\centering
\includegraphics[width=0.95\textwidth]{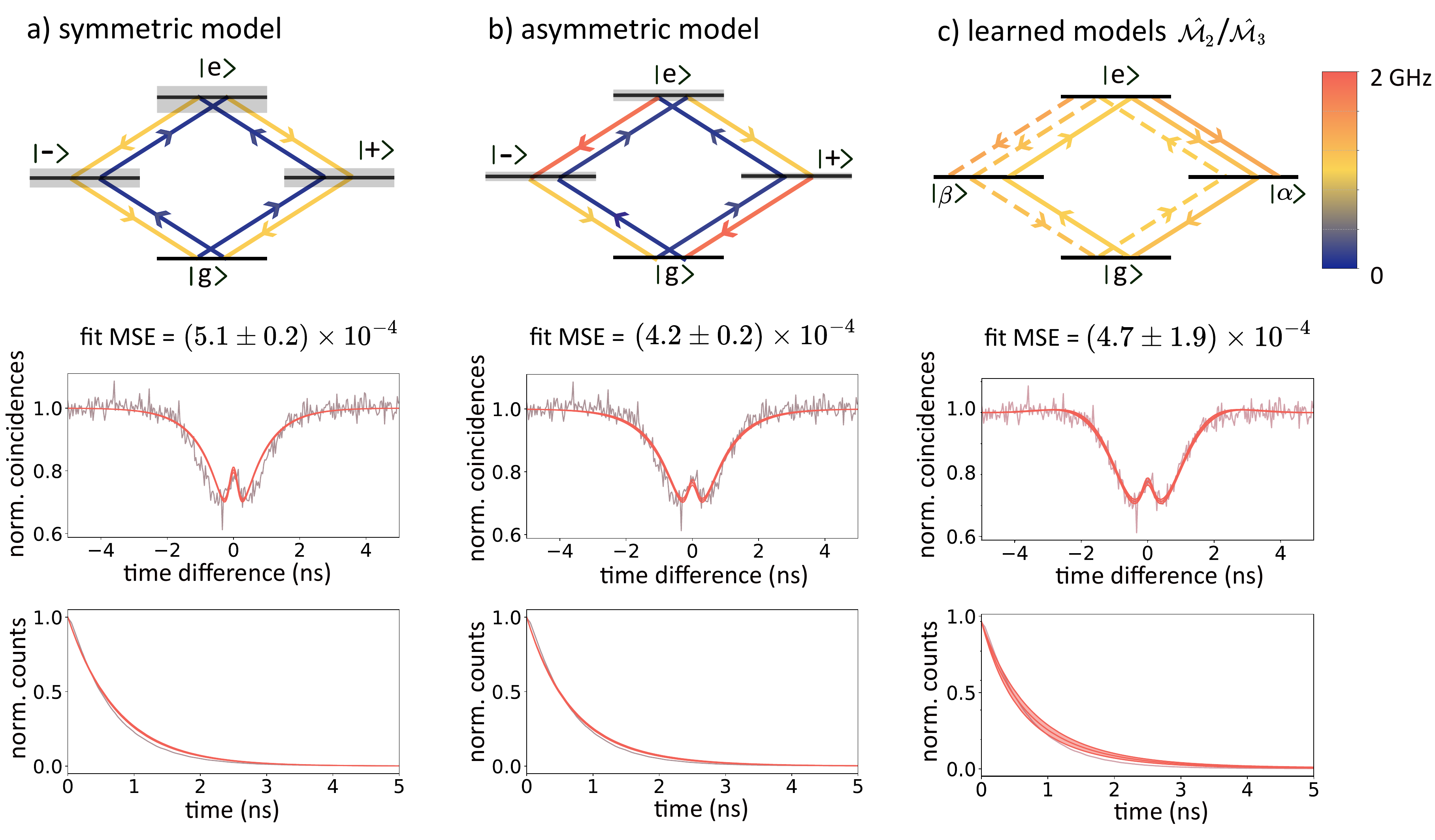}
\caption{ 
\textbf{Comparison between models for the experiments on the two-emitters system.} We comparing three models for the dynamics of the two emitters data (columns \textbf{(a)}, \textbf{(b)}, and \textbf{(c)}). For each model, we plot a visual depiction with arrows representing the relevant Lindbladian processes in the top row, and the corresponding fits of the experimental data (lifetime in the middle row, $g^{(2)} (\tau)$ in the bottom row. In the three sketches in the top row, the rates corresponding to each process are represented with colours, according to the colourbar on the right.  For the fits, the experimental data is shown in gray and simulations from the respective model in red. Model simulations are computed for, respectively, 70000, 19000, 20000 samples, plotting the mean with a shaded line corresponding to one standard deviation. The ``symmetric'' model (\textbf{(a)}) from Ref.~\cite{cygorek_signatures_2023} keeps excitation, decay and dephasing rates identical for both emitters; the ``asymmetric'' model (\textbf{(b)}) allows excitation, decay and dephasing rates to differ between both emitters; the learned models (\textbf{(c)}) corresponds to models $\mathcal {M}_{2A}$ (solid lines) $\mathcal {M}_{3A}$ (dashed lines) from Fig.~4.  Note the MSE value quoted here is for the range of data shown in present panels featuring a reduced range compared to that shown in Fig.~4(d) and 4(e). 
In the schematic model depiction (top row), gray broadening of levels represents energy fluctuations giving rise to pure dephasing. The line thicknesses are representatives of the rate of the respective process with common normalization across all three models. In the probability distribution for the ``asymmetric model'' in (b) we identify four clusters: two with asymmetric rates, and two with symmetric rates very similar to (a): we depict here one of the asymmetric clusters. Each pair of symmetric/asymmetric clusters in (b) only differs by swapped dephasing rates.
}
\label{fig:comparison}
\end{figure*}

In turn, this results in `cooperative emission' without super-radiant rate enhancement. This explanation assumes identical emitters with only pure dephasing as decoherence, characterized by the following dynamical model governing its (generally non-monitored) evolution~\cite{cygorek_signatures_2023}
We proceed by exploring how the model classes $\mathcal{\hat{M}}_{2}$ and $\mathcal{\hat{M}}_{3}$ compare to the original interpretation of the data offered in Refs.~\cite{koong_coherence_2022, cygorek_signatures_2023}. These works introduce an explanation of the $g^{(2)}(\tau=0)$ anti-dip as arising from measurement-induced correlations. This assumes a monitored detection channel that only detects a subset of emitted photons, for whom it cannot  (through the geometrical setup) distinguish from which emitter the photon originated. This corresponds to performing a $(\sigma_1^- + \sigma_2^-)/\sqrt{2}$ projection of the two emitter system into an entangled state \cite{cygorek_signatures_2023}, where the $\sigma_j$ operators for emitter $j$ are defined with respect to the emitter (site) basis~\footnote{To be explicit, we have used the short notation $(\sigma_1^- + \sigma_2^-)/\sqrt{2} \equiv (\sigma_1^- \otimes \mathcal{I}_2 + \mathcal{I}_1 \otimes \sigma_2^-)/\sqrt{2}$.}, rather than the a priori unspecified single excitation manifold states $ \ket{\alpha}$, and $\ket{\beta}$ from above: 
\begin{align}\label{Eq: Moritz's Model}
    \dot{\rho}= &\gamma \left( \mathcal{D}[\sigma^-_1](\rho) +\mathcal{D}[\sigma^-_2](\rho) \right) + \\
    &\gamma_p \left(\mathcal{D}[\sigma^+_1](\rho) +\mathcal{D}[\sigma^+_2] (\rho) \right) \nonumber + \\
    &\gamma_d \left( \mathcal{D}[(\sigma^+_1\sigma^-_1)](\rho) + \mathcal{D}[(\sigma^+_2\sigma^-_2)](\rho) \right) \nonumber.
\end{align}
Here $\mathcal{D}[L](\rho) = L_j \rho L_j^\dagger - \{ L_j^\dagger L_j,\rho\}/2$, $\gamma$ is the optical and $\gamma_p$ the pumping rate,  and $\gamma_d$ the dephasing rate.
Based on Eq.~\eqref{Eq: Moritz's Model} and post-selected symmetric detection events, Ref.~\cite{koong_coherence_2022,cygorek_signatures_2023} establish a qualitative similarity of the measured and calculated $g^{(2)}(\tau)$ trace, but did not provide a quantitative analysis that also fits the lifetime curves.

In Fig.~\ref{fig:comparison}, we present a comparison between the model of Eq.~\eqref{Eq: Moritz's Model} (including its indispensable projection onto a symmetric superposition state for monitored optical decay events) and model $\mathcal{\hat{M}}_2$/$\mathcal{\hat{M}}_3$ found by our algorithm (Fig.~\ref{fig:double_emitter}(a)). In each case, we optimize all model rates to obtain the best fit to the data, and we show fits (red curve) to the experimental lifetime and $g^{(2)}(\tau)$ data (gray curve), together with the MSE of the fit. Fig.~\ref{fig:comparison} (a) illustrates the exact model of Eq.~\eqref{Eq: Moritz's Model} with identical rates for both emitters (labeled as ``symmetric''), while Fig. ~\ref{fig:comparison} (b) is a variant of this model where the values of $\left( \gamma, \gamma_p, \gamma_d \right)$  are allowed to differ between the two emitters (labeled as ``asymmetric'').  Fig.~\ref{fig:comparison} (c) displays  $\mathcal{M}_{2A}$ (solid lines) and $\mathcal{M}_{3A}$ (dashed lines), together with the fits for $\mathcal{M}_{2A}$ (very similar results are obtained for $\mathcal{M}_{3A}$).

The relative weighting between lifetime and intensity correlation data differs between panels (a) and (b) vs (c), with the former weighting $g^{(2)}(\tau)$ vs lifetime data at $75\%:25\%$ but the latter placing a much greater ($>99.9\%$) importance on the $g^{(2)}(\tau)$ as this benefited the broader exploration of the model space during the preceding learning stage. This is reflected in our learned model showing the best intensity correlation agreement at the expense of a poorer fit and larger uncertainty for the associated lifetime trace. Indeed, as a consequence, panel (c) shows by far the closest agreement between data and fit around the shoulders of the $g^{(2)} (\tau)$ function and across the central dip~\footnote{It might be interesting to explore dynamic or adaptive relative weighting of different types of input data as a means to enable both broad exploration and then optimisation in separate stages}.
Comparing overall MSE values, our optimized learned model narrowly beats the symmetric model of Eq.~\eqref{Eq: Moritz's Model}, and does so not only with fewer Lindblad terms but also with a more balanced relationship between excitation and emission rates. Conversely, the `asymmetric' model (panel (b)) achieves the best combined MSE, however, it involves the largest number of independent processes and rates combined with the largest difference across the values for relevant rates, which might raise concerns about over-fitting. 
A noteworthy difference between panels (a,b) and (c) is that our learned model does not require a pure dephasing term, which was, however, reported as being integral to matching the shape of the data in Refs.~\cite{koong_coherence_2022, cygorek_signatures_2023}. 

\section {Discussion}
We have presented a Bayesian algorithm to automatically find (minimal) GLSK models that are consistent with measured data. We have demonstrated the application of the method to a set of case studies relevant for quantum optics, such as a single quantum emitter under coherent optical excitation, and cooperative emission for a pair of quantum emitters. We show that the algorithm can identify a ranked list of model classes, with each class including a set of physical processes (e.g. optical excitation, photon emission, dephasing) and the corresponding rates. In particular, the application to the case of two emitters finds a possible model that, to the best of our knowledge, was not previously discussed in the quantum optics literature, and that appears to fit the experimental data better than models currently available.

Our algorithm appears to identify as most likely explanation for the observed experimental results two models alternative to the physically motivated one of Eq.~\eqref{Eq: Moritz's Model}. This is likely the result of the penalization we impose on increasing number of terms (Eq. ~\ref{eq:model_prior}), to minimize the risk of over-fitting, which favor models that can fit the data with fewer processes. On the one hand, our learned model achieves a similarly good, or better, fit to the data with fewer terms, and does so with a more balanced relationship between contributing rates. 
On the other hand, both our learned models as well as that of Ref.~\cite{koong_coherence_2022} involve non-monitored decay channels: however, in the latter case only a small minority of post-selected detection events is assumed to constitute the measured signal, and our implementation includes the requisite symmetric projection as a manual intervention, a step that is not available to the learning algorithm. Finally, our learning algorithm is under constraints of distributing the total transition dipole moment of two individual emitters consistently with basis choice and across the different excitation manifolds, equipping it with additional leeway over a microscopically founded model.

Seeing our learning algorithm accomplishes its objective of identifying viable and minimal physical models, an important question for future development is how to combine such explorative learning with a larger degree of input about known physical properties and constraints. The choice of the best penalization on increasing number of terms is an open question: while this is important to avoid over-fitting, the strategy to identify the most suitable penalization function and optimal values for the corresponding meta-parameters is not obvious and has a strong impact on the ranking of the learned models~\footnote{Note that our choice of penalising a large number of terms is in spirit related to the idea of incentivising a low rank for the Kossakowski matrix, however, our choice of operator basis means there is no direct correspondence between that rank and the number of terms in our model.}.

This work highlights the power of computer algorithms to suggest alternative models that researchers may not have considered when designing their experiments. In particular, our algorithm can help clarifying to what degree experiments do validate a given physical model or whether other alternatives have not been fully ruled out. The use of Bayesian inference enables us to quantify uncertainty in both the processes and the corresponding rates, mitigate over-fitting, and to address multi-modality, with the possible existence of multiple models that can explain the experimental data. The algorithm can be applied to a variety of practical situations, from continuously monitored quantum systems \cite{gammelmark_bayesian_2013}, to the benchmarking of quantum circuits \cite{samach_lindblad_2022}, to quantum sensing \cite{rosenberg_witnessing_2025}.

A possible direction of future research is investigating how our model identification approach could be expanded into a computational framework that additionally suggests optimal experiments to discriminate between different model classes. This would generalize current experiments on real-time adaptive optimization of quantum control sequences for systems with known models \cite{bonato_optimized_2016, arshad_real-time_2024, berritta_physics-informed_2024}, and it could for example be based on model-aware reinforcement learning  \cite{belliardo_application_2024} or deep Bayesian experimental design \cite{sarra_deep_2023}.

As with any classical algorithm processing quantum systems, complexity scales exponentially with system size, limiting the current applicability to quantum systems of relatively  small size. Having access to observables that are closer to a tomographic set than the case we have looked at is expected to lead to improved performance, as is the ability to restrict the number of noise processes under consideration. In any case, future availability of a trusted digital or analog quantum simulator may aid the calculation of the likelihood \cite{schlimgen_quantum_2022, matthew_pocrnic_quantum_2023, moss_enhancing_2023, luo_variational_2024, ding_simulating_2024, olaya-agudelo_simulating_2024, trivedi_quantum_2024}.

A further route to improving performance is the Markov Chain Monte Carlo algorithm itself. A possible route is to leverage partial gradient-based optimization with  a differentiable solver~\cite{Craig2024}.
Recently, researchers have also presented a quantum Markov chain Monte Carlo algorithm \cite{david_layden_quantum-enhanced_2023} that converges to the correct distribution in fewer iterations than common classical MCMC alternatives, with a polynomial speedup between cubic and quartic in simulations. Such quantum MCMC additionally suggests unusual robustness to noise, as experiments on current noisy intermediate quantum (NISQ) devices  found it to converge to the correct distribution slower than in simulations under ideal conditions, but still faster than common classical alternatives. 

\vspace{0.5cm}
\textit{Note} - While this manuscript was under review, we became aware of an independent work \cite{fioroni_learning_2025} with similar goals but a different algorithm and application case.

\section*{Acknowledgements}
We thank Zhe Xian Koong, Raffaele Santagati, Antonio Andrea Gentile, Brian Flynn, Nicholas Werren, Federico Belliardo, Dorian Gouzou and Moritz Cygorek for helpful discussions. This work is funded by the Engineering and Physical Sciences Research Council (EP/S000550/1 and EP/V053779/1, and through the UK Quantum Technology Hub in Quantum Imaging EP/T00097X/1 and the Integrated Quantum Networks Hub EP/Z533208/1), the Leverhulme Trust (RPG-2019-388 and RPG-2022-335) and the European Commission (QuanTELCO, grant agreement No 862721; QuSPARC, grant agreement No 101186889). 

\section*{Author Contributions}
CB and EMG conceived the project. SW, YA, EMG and CB designed the algorithm. SW implemented the code and generated the learning results. All co-authors analyzed the generated data and wrote the manuscript.

\appendix
\section {Lindblad open system modelling}
\label{app:lindblad}

Our objective is to find models $\mathcal{M}$ that describe the dynamics of the system through Eq.~\eqref{eq:lindblad}. We summarize here the main terminology used throughout the paper:
\begin{itemize}
    \item \textbf{Term: } A single matrix from the matrix basis, with which processes are constructed (terms do not contain addition of multiple matrices).
    \item \textbf{Process: } A matrix representing either a single Lindblad operator ${L}_s$ or a single linear component of the Hamiltonian ${H}_r$.
    \item \textbf{Rate: } Real numerical coefficient ($\omega_h$, $\gamma_l$) associated with each process. We here explicitly include frequencies of Hamiltonian terms as `generalized rates' and note that both have the same units as inverse time when $\hbar=1$.
    \item \textbf{Model: } A set $\mathcal{M} = \mathcal{M}(\boldsymbol{\Theta}; \boldsymbol{\theta})$ of `rates' and processes, defined as
    \begin{align}
    \mathcal{M} &= \mathcal{M}(\boldsymbol{\Theta}; \boldsymbol{\theta}) \nonumber \\
    & = \mathcal{M}(\{ {H}_h,h = 1...n_h;{L}_l,l=1...n_l\}; \\
    & \qquad \{\omega_h,h = 1...n_h;\gamma_l,l=1...n_l,b\}). \nonumber 
    \label{Model}
\end{align}

\end{itemize}

The Lindblad master equation~\eqref{eq:lindblad} can be straightforwardly mapped to a Liouvillian superoperator $\mathcal{L}$ \cite{fabrizio_minganti_spectral_2018}, resulting in the compact form
\begin{equation}
    \dot{\rho}(t) =\mathcal{L} \rho(t)
\end{equation}
with the formal solution $\rho(t) = e^{\mathcal{L}t}\rho(0)$, where $\rho(0)$ is the initial state of the system.
For a $d$-dimensional system, the Liouvillian has a representation as a $d^2 \times d^2$ matrix with the density operator $\rho(t)$ written in vectorized form as a vector of length $d^2$ (in the main text denoted by $\vert \rho(t) \rangle$). This Liouvillian matrix features complex-valued elements and generally does not have full rank. Its null-space~\footnote{It may also have purely imaginary eigenvalues representing a limit cycle instead of a steady state, albeit not in the context we consider here.} defines the steady state(s) of the system.

A practical advantage of generating dynamics from a Liouvillian is that $\exp{\mathcal{L}t}$ only depends on time in a parametric fashion (analogous to the time evolution operator). This means no integration of (a set of coupled ordinary) differential equations (as in Eq.~\eqref{eq:lindblad} is required for evolving the quantum state, and this also helps with finding steady states as the eigenvector(s) associated with (the) zero-valued eigenvalue(s).

As discussed in the main text, we thus express our models as Lindblad master equations for easy interpretability and guaranteed physicality, but then map these master equations to their corresponding Liouvillian representations \cite{fabrizio_minganti_spectral_2018} for generating dynamics and determining steady states.

\section {Defining the operator space}
\label{app:operator_space}

\subsection{Constructing a library of possible operators}\label{SM: Operator Space}

Our approach utilizes master equations for representing models, whilst switching to the Liouville representation generating dynamics. For a $d$-dimensional system, the former feature one Hamiltonian and a collection of Lindblad operators that are each $d \times d = d^2$ matrices. By contrast, the Liouvillian superoperator has $d^4$ (complex) elements. The physical system considered in our work (two quantum dots) is described by a four-level system, so that $d^2=16$ and $d^4=256$.

To assemble Hamiltonian and Lindblad operators, we require an operator basis. The canonical operator basis for our effective four level system features $16$ elements~\footnote{The $d=4$ generalized Gell Mann matrices are an alternative equivalent choice}. Seeing we generally have complex elements, this would lead to up to 32 `expansion coefficients' per (Lindblad) operator (and reduced by the constraint of Hermiticity for the Hamiltonian). The non-linear nature of the Lindblad dissipator in Eq.~\eqref{eq:lindblad} then results in an excessively large number of coefficients to be determined for each model candidate (not least since the number of required Lindblad operators is unknown a priori).

We wish to identify models that are simultaneously physically interpretable and maximally `sparse', i.e.~defined by a minimal number of associated `rates'. For this reason we consider a more restricted physically motivated basis set and procedure to construct operators. As detailed in the main text, we instead only allow operators comprised of a maximum of two equally weighted basis elements, resulting in 105 operator matrices for the two emitter learning, and only 10 (including an inconsequential but present identity) for the single emitter. Each of these, if present in a model, will have an associated generalized rate characterized by a positive real number.

\
\subsection{Identification of Emission and Excitation Operators} \label{app: Emission and Excitation Detection}
As quantum optics experiments involve the optical excitation of quantum systems and the detection of emitted photons,  the corresponding models need to identify excitation and emission events. This is facilitated by the Lindblad master equation approach, as compared to the  Liouvillian framework, given that each operator has a direct physical interpretation. In general, the excitation operators are known, as they represent the action of the experimenter on the system.

Regarding photon emission and detection, any experimental setup has some selectivity in terms of optical wavelength range, polarisation and spatial modes so that not all allowed optical transitions necessarily lead to an experimental photo-detection events. This selectivity needs to be included in the learning process as part of the prior information. In the following, we group the relaxation operators leading to a photon detection event in the class $L_\downarrow$.

\begin{widetext}
\begin{algorithm}

    \caption{Linblad Master Equation Learning - Algorithm Overview}
    \label{Alg:MCMC}
        $\{\boldsymbol{y}_k\}$                    \tcp*{$\{\boldsymbol{y}_k\} \equiv$ Set of Experimental Measurements}
        $N$                      \tcp*{$N \equiv$ Number of Steps} 
        $n_{\textrm{start}} = n_l + n_h$  \tcp*{$n_{start} \equiv$ Guessed Number of Processes}
        Initialize $\mathcal{M}_0 =  \mathcal{M}(\Theta = \{\}, \theta = \{\})$; 
        
        \For{j $\in$ 1 to $n_{\textrm{start}}$}{
            $\text{M} \sim \{M\}_{add}$ \tcp*{$\{M\}_{add} \equiv$ Moves That Add Processes}
            $\mathcal{M}_0 \leftarrow \text{M}[\mathcal{M}_0]$;
        }
        Sample $\beta_k \sim \Gamma(a_k,b_k)$ \tcp*{$a_k$ and $b_k$ set priors}
        Compute $Pr(\mathcal{M}_0,\beta_k|\{\boldsymbol{y}_k\})$ 
        
        \For{i $\in$ 1 to $N_{\textrm{iter}}$}{
            Generate candidate $\mathcal{M}'|\mathcal{M}_{i-1}$ and compute $Pr(\mathcal{M'},\beta_k|\{\boldsymbol{y}_k\})$; 
            
            Compute accept/reject ratio $\rho(\mathcal{M}_{i-1},\mathcal{M}')$; 
            
            Sample $R \sim \mathcal{U}[0,1]$;\tcp*{From a uniform distribution}
            
            \eIf{$R\leq \alpha$}{
                $\mathcal{M}_i \leftarrow \mathcal{M}'$  \tcp*{Accept}
                
                $Pr(\mathcal{M}_i,\beta_k|\{\boldsymbol{y}_k\}) \leftarrow Pr(\mathcal{M'},\beta_k|\{\boldsymbol{y}_k\})$;
            }{
                $\mathcal{M}_i \leftarrow \mathcal{M}_{i-1}$ \tcp*{Reject}
            
                $Pr(\mathcal{M}_i,\beta_k|\{\boldsymbol{y}_k\}) \leftarrow Pr(\mathcal{M}_{i-1},\beta_k|\{\boldsymbol{y}_k\})$;
            }
               Sample $\{\beta_k\}_k$ using Gibbs sampling;
            }
\end{algorithm}
\end{widetext}

\section{Algorithm Overview} \label{app:algorithm_overview}

Our computational approach, sketched in Fig~\ref{fig:algorithmic figure} and described by the pseudo-code in Algorithm 1, constructs Markov chains of models $\lbrace \mathcal{M}_i \rbrace$. At the end of this process, the models generated are then clustered into model classes $\mathcal{\hat{M}}_q$ and subsequently ranked based on how frequently they have been generated.  At each iteration of the Markov chain, a model is proposed from the most recently generated model, via bespoke so-called moves. This candidate model is then accepted or rejected according to the standard acceptance rate for Metropolis-Hastings algorithms. 

The algorithm takes as input parameters the experimental datasets $\lbrace \mathbf{y}_k \rbrace$, the number of iterations $N_{\textrm{iter}}$ and the expected number of Hamiltonian ($n_h$) and Lindbladian ($n_l$) processes. A starting model $\mathcal{M}_0$, consisting of $n_h$ Hamiltonian and $n_l$ Lindbladian processes is then initialised randomly.

\subsection{Bayesian inference}
Following the notation introduced above, the proposed Monte Carlo sampler targets the model posterior distribution $Pr(\mathcal{M}(\boldsymbol{\Theta},\boldsymbol{\theta}),\boldsymbol{\beta}|\{\textbf{y}_k\})$, where the vector $\textbf{y}_k$ represents the experimental data from the $k$th experiment and where $\boldsymbol{\beta}=\{\beta_k\}$ are the (inverse) noise levels affecting the different experiments. Here we consider only two experiments ( lifetime ($k = $`LT') and $g^{(2)} (\tau)$ ($k = $ `$g^{(2)}$') measurements) but the method applies to more complex experimental setups.
 The posterior distribution for $(\mathcal{M}(\boldsymbol{\Theta},\boldsymbol{\theta}),\boldsymbol{\beta})$ is obtained using Bayes rule:
 
\begin{align} 
\label{eq: bayes appendix}
& Pr(\mathcal{M}(\boldsymbol{\Theta},\boldsymbol{\theta}),\boldsymbol{\beta})|\{\textbf{y}_k\}) = \nonumber \\
& \qquad \propto \lambda(\{\textbf{y}_k\}_k | \mathcal{M}(\boldsymbol{\Theta},\boldsymbol{\theta}),\boldsymbol{\beta}) p(\mathcal{\boldsymbol{\Theta}},\boldsymbol{\theta},\boldsymbol{\beta}), 
\end{align}

where $p(\boldsymbol{\Theta},\boldsymbol{\theta},\boldsymbol{\beta})=p(\boldsymbol{\Theta})p(\boldsymbol{\theta}|\boldsymbol{\Theta})p(\boldsymbol{\beta})$ is a user-defined joint prior distribution and $\lambda(\{\textbf{y}_k\} | \mathcal{M}(\boldsymbol{\Theta},\boldsymbol{\theta}),\boldsymbol{\beta})$ is the likelihood given by:

\begin{align}
\label{eq: likelihood appendix}
&\log(\lambda(\{\textbf{y}_k\}_k |\mathcal{M}(\boldsymbol{\Theta},\boldsymbol{\theta}),\boldsymbol{\beta}))  = \nonumber \\
& \qquad = \sum_{k \in \{LT , g^{(2)}\}} \frac{-\beta_k}{2}\boldsymbol{\Delta y}_k^T\boldsymbol{W}_k(\boldsymbol{\tau}_k)\boldsymbol{\Delta y}_k
\end{align}
with 
\begin{equation}
\label{eq:delta_y_k}
    \boldsymbol{\Delta y}_k = \boldsymbol{y}_k (\boldsymbol{\tau}_k|\mathcal{M}(\boldsymbol{\Theta},\boldsymbol{\theta})) - \textbf{y}_k (\boldsymbol{\tau}_k).
\end{equation}
Here the vector $\boldsymbol{\tau}_k$ encodes the photon arrival times for the $k$-th experiment, and $\boldsymbol{W}_k(\boldsymbol{\tau}_k)$ is a weighting function that controls the relative importance of each experimental data point within $\boldsymbol{y}_k (\boldsymbol{\tau}_k)$. 

\subsection{Prior models and data weighting} 
The parameters $\{\beta_k\}$ are assigned independent Gamma prior distributions with mean $m_k$ and variance $v_k$, which are set from preliminary data simulations (to analysis the amount of random noise in the data). In practice, since we wish to better fit the $g^{(2)}$ data than the lifetime data (which contain less information), we set the mean $m_{g^{(2)}}$ slightly higher than the expected noise precision.

Although the number of the Hamiltonian and Lindbladian operators is unknown, one may want to favor simpler models, composed of few processes, to avoid overfitting. This is achieved here by assigning a prior model to the Hamiltonian and Lindbladian operators, but also by assigning priors to the associated rates. To promote simple models, we first use the following prior 
\begin{eqnarray}
\label{eq:model_prior}
    p_m(\boldsymbol{\Theta}) &=&
    \left(\binom{len(\boldsymbol{\bar{H}})}{n_h}
    {\binom{len(\boldsymbol{\bar{L}})}{\boldsymbol{L}^{(a)}_i} {\binom{len(\boldsymbol{\bar{L}})}{n_l}}}\right)^{-1}\nonumber\\
    &\times& e^{\frac{-n_h}{\eta_{H}}}
    e^{\frac{-n_l}{\eta_{L}}}
     e^\frac{-n_c}{\eta_c},
\end{eqnarray}
where $n_h$ denotes the number of Hamiltonian processes, $n_l$ denotes the number of Lindbladian processes, and $n_c$ is the number of terms involved in a given Lindbladian process. Those values in Eq. \eqref{eq:model_prior}, are penalized by the tuning parameters $\eta_H$, $\eta_L$ and $\eta_c$, respectively. Moreover, $\boldsymbol{\bar{H}}$ and $\boldsymbol{\bar{L}}$ denote the full sets of admissible Hamiltonian and Lindbladian operators, respectively, and they are introduced to make sure our prior is properly normalized.

Finally, we define a prior model $p(\boldsymbol{\theta}|\boldsymbol{\Theta})$ for the rates and energies, conditioned on $\boldsymbol{\Theta}$. While it is possible to tailor priors for each operator, here we use identical Gamma priors for all rates (mean 0.6 GHz, standard deviation 12 GHz) which ensures positivity,  promotes rates and energies with similar magnitude (in line with preliminary data investigation, whereby we expect dynamics at the nanosecond timescale), and prevents severe over-fitting by penalizing small rates/energies.

\subsection {Computing the likelihood of models}
The application of Bayesian inference requires the computation of the expected experimental result $\boldsymbol{y}_k (\boldsymbol{\tau}_n|\mathcal{M}(\boldsymbol{\Theta},\boldsymbol{\theta}))$ for a given model $\mathcal{M}(\boldsymbol{\Theta},\boldsymbol{\theta})$. To provide a link with the experiments, which include optical excitation of the system and the detection of the emitted photons, we identify two subsets of processes within $\Theta$, corresponding to excitation processes ($L_\uparrow$) and relaxation processes ($L_\downarrow$, denoting any operator that triggers a detector click).

In the calculation of the expected result for the lifetime measurement, we consider the decay of the system from a fully excited state, taking the sum of projective measurements for each relaxation process. This results in :

\begin{equation}
\label{Eq: LT Calculation}
    \boldsymbol{y}_{LT} (\boldsymbol{\tau}|\mathcal{M}_i) = \sum_{v \in L_\downarrow} Tr(v^\dagger V(\tau) \rho_{e} v)  \,,
\end{equation}
where the system density matrix $\rho_C$ is evolved using a Liouvillian with no excitation processes $V(\tau) = e^{\mathcal{L}_\downarrow \tau}$ (based on the specific model $\mathcal{M}(\boldsymbol{\Theta},\boldsymbol{\theta})$) and projective measurements for every operator in $L_{\downarrow}$. 

The $g^{(2)}(\tau)$ involves computing the correlation between clicks of two detectors.  In the case of multiple emission processes, one needs to take into account the combinatorial set of  processes being interrogated:

\begin{equation}
\label{Eq:g2_Calculation}
    \boldsymbol{y}_{g^{(2)}} (\boldsymbol{\tau}|\mathcal{M}_i) = \frac{1}{\boldsymbol{y}_{g^{(2)}} (\boldsymbol{\infty}|\mathcal{M}_i)} \sum_{(v,u) \in L_\downarrow} Tr[vv^\dagger V(\tau)u^\dagger\rho_{ss}u ]
\end{equation}

From the ideal likelihoods in Eq.~\eqref{Eq: LT Calculation} and~\eqref{Eq:g2_Calculation}, we further include experimental imperfections, convolving with the temporal response function of the photon detectors and the background dark count rate and adding the background dark count rate. More details about the computation of the likelihood are described in Appendix \ref{sec:likelihood}.

\subsection{Model-updating moves}
At each iteration of the algorithm, a new model is created and assessed using simultaneously 1) its fit to the experimental data and 2) its agreement with our a priori knowledge. A new model is created through the application of one model-updating move, randomly chosen from a library of moves, briefly summarized here:

\begin {itemize}
\item \textit{rate update}: one of the rates, randomly chosen, is resampled (from a Gamma distribution);
\item \textit{process birth}: a new operator (Hamiltonian or Lindbladian) is added to the model;
\item \textit{process death}: an operator (Hamiltonian or Lindbladian) is randomly selected from the model and removed;
\item \textit{process swap}: a Hamiltonian (Lindbladian) operator is randomly selected from the model and replaced by a different Hamiltonian (Lindbladian) operator;
\item \textit{conjugate Lindbladian process birth}: the complex conjugate of an existing Lindbladian operator is added (e.g. a photon emission operator, when an absorption operator is found); 
\item \textit{conjugate Lindbladian process death}: the complex conjugate of an existing Lindbladian operator is removed;
\item \textit{conjugate Lindbladian process swap}: selects a pair of Lindblad operators that are Hermitian conjugate of one another and swaps them out for another pair of Hermitian-conjugate Lindblad operators that do not already exist in the model;

\end{itemize}

The ``conjugate Lindbladian'' moves only apply to Lindbladian operators as Hamiltonians are Hermitian.

\subsection {Accept/reject procedure.} 
Let's denote by $\bfx=(\mathcal{M}(\boldsymbol{\Theta},\boldsymbol{\theta}),\boldsymbol{\beta})$ the current state of the Markov chain and by $\bfx^*=(\mathcal{M}(\boldsymbol{\Theta}^*,\boldsymbol{\theta}^*),\boldsymbol{\beta}^*)$ a proposed candidate model.  According to the Metropolis-Hastings accept/reject ratio, the candidate is accepted with probability $a(\bfx,\bfx^*)=\textrm{min}\left(1,\rho\right)$, with 
\begin{equation} \label{Eq: accept reject}
    \rho = \frac{Pr(\bfx^*|\{\textbf{y}_k\})P(\bfx^*,\bfx)}{Pr(\bfx|\{\textbf{y}_k\})P(\bfx,\bfx^*)},
\end{equation}
where $P(\bfx,\bfx^*)$ denotes the probability for proposing $\bfx^*$ from $\bfx$, and $P(\bfx^*,\bfx)$ the probability of the reverse move from $\bfx^*$ to $\bfx$.

\subsection{Creating a ranked list of model classes.}
\label {app:ranked_list}
Several Lindblad master equations in the Markov chains may give rise to similar behavior, and indeed multiple distinct master equations may even possess identical Liouvillian representations. It is therefore instructive to classify the learnt models into ``Model Classes'' ($\mathcal{\hat{M}}_q$) that express similar physics. To reliably expose differences in model behaviour, we convert master equations to Liouvillian operators (uniformly sampling $10\%$ to reduce dataset size while preserving statistics). The vectorized Liouvillian matrices are projected using Principal Component Analysis (PCA) \cite{Pearson1901} and clustered using k-means \cite{Lloyd1982}, choosing the number of clusters with the ``elbow method'' \cite{david_j_ketchen_application_1996}. 

After identifying the clusters, we analyse them separately. In Fig 3(b) and 4(b), we rank the model classes by popularity, computing the number of times a model belonging to each class appears in the set of chains. Furthermore, we also consider how well each model class fits the experimental data, computing the mean squared error between simulated datasets with parameters from the learned model class and the actual experimental data (Fig. 3(a) and Fig 4(a)).

\section{Additional details on computing the likelihoods}
\label{sec:likelihood}

Here we provide more details on how the likelihood functions \eqref{Eq: LT Calculation} and \eqref{Eq:g2_Calculation}, and discuss how experimental imperfections are included.

\subsection{Lifetime simulation }
\label{ref:lifetime_likelihood}
A lifetime experiment consists of exciting the system with a short pulse (shorter than the system relaxation dynamics) to initialize it in its excited state, and timing the arrival of subsequently the emitted photons at the photo-detector. It thus provides a means of tracking the excited state population over time.

We begin with a simple two-level system, for which we can write
\begin{equation*}
     \boldsymbol{y}_{LT} ( \boldsymbol{\tau}) = Tr \left[ \sigma_+\sigma_- \rho({ \boldsymbol\tau}) \right],
\end{equation*}
where $\rho(t)$ is the density matrix describing the state of the two-level system a time $t$ after the initial excitation. The initial state is set as $\rho(0) = \ket{e}\bra{e}$ with $\ket{e}$ being the excited state of the two-level system. We evolve $\rho(t)$ according to $\rho (t|\mathcal{M}_i) = e^{\mathcal{L} (\mathcal{M}_i) \, t} \rho (0)$, based on the model $\mathcal{M}_i$ being considered, so that the lifetime observation model is given by: 
\begin{equation*}
    \boldsymbol{y}_{LT} (\boldsymbol{\tau}|\mathcal{M}_i) = Tr \left[ \sigma_+ \sigma_- e^{\mathcal{L} (\mathcal{M}_i) \boldsymbol{\tau}}\rho(0) \right] .
\end{equation*}

A system with more than two levels may have the additional complexity of an interplay between different transition rates, where only some decay processes result in detectable optical emission. Furthermore, the specific experimental setup may add some wavelength, polarization or spatial mode selection on the optical emission, so that only a subset of the emitted photons do actually result in a detector click. We identify the subset of the detected modes as $\lbrace L_{\downarrow}\rbrace$, such that the corresponding lifetime observation model can be written as Eq. \eqref{Eq. LT Calculation}.

As described in Eq. \eqref{eq: likelihood appendix},  weights are assigned to the different experiments, and to give more importance to some data points. For the lifetime measurements, we use a weighting based on a normalized modulus of the derivative of the data:
\begin{equation}
    \boldsymbol{W}_{LT}(\boldsymbol{\tau}) = \left|\frac{d\boldsymbol{y}_{LT}(\boldsymbol{\tau})}{d\boldsymbol{\tau}}\right| \frac{1}{N},
\end{equation}
where $N = \sum_\tau \boldsymbol{y}_{LT}(\boldsymbol{\tau})$. This choice privileges data points exhibiting higher local variations, since those points are more informative than regions where the signal is flat.

\subsection{Intensity correlation simulation}
\label{ref:g2_likelihood}

The second order intensity correlation, $g^{(2)} (\tau)$, characterizes the relative conditional probability that a second (detected) photon emission event will occur some time $\tau$ after a first photon triggered a detector. The usual definition of the $g^{(2)} (\tau)$ for a two-level emitter with optical decay operator $v$ is given by
\begin{equation}
    g^{(2)}(\tau) = \frac{\langle v^\dagger(0)v^\dagger(\tau)v(\tau)v(0)\rangle_{{ss}} }{\langle v^\dagger(0)v(0)\rangle_{{ss}}\langle v^\dagger(\tau)v(\tau)\rangle_{{ss}} }, 
\end{equation}
once we relate the detected photon mode(s) back to the transition operator of the emitter, and where $\langle\bullet\rangle_{ss}$ indicates the expectation value with respect to the steady state density operator of the system, $\rho_{ss}$, under cw driving (we do not consider pulsed excitation for these type of measurement in this work). 

Considering the un-normalized form $G^{(2)} (\tau)$ and making use of the quantum regression theorem \cite{breuer_theory_2007}, we obtain
\begin{equation*}
    G^{(2)}(\tau) = Tr \left[ v (V(\tau) \underline{\rho}) v^\dagger \right] ,
\end{equation*}
where $V(t) = e^{\mathcal{L}(\mathcal{M}_i) t}$ propagates the un-normalized pseudo-density matrix $\underline{\rho} = v\rho_{ss}v^\dagger$ from the first to the second detection event. We may obtain $g^{(2)}$ by normalizing such that $G^{(2)}(\tau \to \infty) = 1$. In practice, we need to take $\tau$ beyond any correlation / coherene timescale of the system, in our case $\tau \gtrsim 5$ ns is sufficient.

Generalizing beyond a single optical decay operator to an extended set of emission processes $L_\downarrow$ we allow the initial operator and the second emission operator to be different:
\begin{equation}\label{Eq: g2 Calculation SI-2}
    G^{(2)}(\tau) = \sum_{(v,u) \in L_\downarrow} Tr \left[ v^\dagger v (V(\tau)u\rho_{ss}u^\dagger) \right] .
\end{equation}
The sum over each combination of the operators in $L_\downarrow$ reflects the fact that our detectors cannot know which specific process generated a particular photon. Upon normalization, we arrive at the expression in Eq. \ref{Eq:g2_Calculation}.

To guide our algorithm and aid discrimination between models featuring a ``dip'' or ``anti-dip'' around $\tau = 0$, we use a Gaussian weighting profile $\boldsymbol{W}_{g^{(2)}} (\boldsymbol{\tau})$ centered at $\tau = 0$.  This gives more importance to the region near $\tau \sim 0$ in the $g^{(2)} (\tau)$ datasets. 

\subsection{Including experimental imperfections}
\label{sec:exp_imperfection}

Compared to the ideal calculations described above, the experimental data is affected by noise and imperfections that must carefully be taken into account for the Bayesian learning procedure to be successful. There are two main sources of imperfections in our experimental data: the instrument response function (IRF) and the dark count rate of the single-photon detectors.

The detector IRF describes the experimental limit in the temporal resolution of photon-detection, related to electronic jitter of the device and the associated electronics. Its effect can be modelled by convoluting the ideal likelihood with the IRF. Based on the specifications of the superconducting single photon detectors used in the experiments, we take the IRF to be a Gaussian function with a full width, half maximum (FWHM) of 240 ps \cite{koong_coherence_2022}.

The background count rate $\mathcal{B}$ describes the rate at which the detector clicks in the absence of incoming photons, due to internal physical processes in the physics of the detector device. The background rate $\mathcal{B}$ is used as a fit parameter in all learning runs.

\section{Meta-parameter Optimization}\label{Sec. Meta-parameters}
The algorithm includes a set of meta-parameters that govern its operation, such as the probability to select a given move, or the complexity of the operator library. Meta-parameters are defined by the user, and they can affect the performance of the algorithm in terms of speed and reliability in exploring the model space. Optimal values will generally depend on the specific problem and experimental data available, and their (at least coarse) optimization needs to be performed on a case by case basis. Meta-parameters have been optimized through extensive numerical simulations, given target performance for criteria such as a large variety of models, small models, ability to fit the experiments, interpretability, acceptance rates, or mixing properties.

\begin{table}
\begin{center}
\begin{tabular}{ |c|c|c| }
\hline
Name & Symbol & Value\\
\hline
Complexity & $\mathcal{C}$ & 2 \\
Proposal Variance & $\sigma^2$ & 0.3\\ 
Lindbladian penalty scaling & $\eta_L$ &  5 \\
Hamiltonian penalty scaling & $\eta_L$ & 2 \\
Linear-combinations penalty scaling & $\eta_L$ & 1 \\
Chain Steps & $N_C $ & 100000 \\
\hline 
\end{tabular}
\end{center}
\caption {Values for the meta-parameters used in the application of our algorithm to learning models for single and coupled quantum emitters, as discussed in the main text.}
\label{table:metaparameters}
\end{table}

In the following, we list the meta parameters that control the likelihood and model space. The specific values chosen for the learning performed in the main text are reported in Table \ref{table:metaparameters}.

\begin{itemize}
    \item \textbf{Proposal Variance ($\sigma^2$)} describes the variance of the Gaussian distribution from which the parameter estimation move samples from. If this value is too large, the algorithm will miss the finer details of the likelihood landscape and take more time to optimize the rates. Conversely, if this parameter is set too small, it will take longer for the rate to converge to a near-optimal value. For our application, given that rates do not to differ beyond an order of magnitude, we found a value of $0.3$ to be an appropriate trade-off.

    \item \textbf{Complexity ($\mathcal {C}$).} As discussed in \ref{SM: Operator Space}, this meta-parameter controls the maximum number of matrices that can be combined in the learning. Human interpretability privileges smaller complexity, however fewer operators may be needed to fit the experimental data if this meta-parameter were set to a larger value. Throughout this work, we chose $\mathcal{C} =2$.

    \item \textbf{Lindbladian and Hamiltonian penalty scaling ($\theta_L$ and $\theta_H$} $\eta_L$ and $\eta_H$) define a user-controlled preference on the number of Lindblad and Hamiltonian terms, respectively, to expect. There is no upper bound on the number of terms, but fewer terms are preferred. In our experiments, we set $\theta_L=\theta_H=2$. 

    \item \textbf{Linear-combinations penalty scaling ($\theta_C$} $\eta_C$) controls the preference on the number of linear combinations of operators to expected in the model. There is no upper bound on the number of terms, but fewer terms are preferred. In our experiments, we set $\theta_C=2$. 

\item \textbf{Chain Steps ($N$)} is the length of the chains the algorithm builds. Though it is important to parallelize learning, a long burn-in might be required to let the algorithm reach the region of high density of the posterior distribution. This parameter needs to be adapted for each dataset.

\end{itemize}

\begin{table}
\begin{center}
\begin{tabular}{ |c|c|c| }
\hline
Move & Tag & Probability (\%)\\
\hline
Rate Update & \textit{RATE} & 40\\
Hamiltonian birth & \textit{BIRTH [H]} & 4\\
Hamiltonian death & \textit{DEATH [H]} & 4\\
Swap Hamiltonian & \textit{SWAP [H]} & 8\\
Lindbladian birth & \textit{BIRTH [L]} & 8\\
Lindbladian death & \textit{DEATH [L]} & 8\\
Swap Lindbladian & \textit{SWAP [L]} & 16\\
Conjugate Lindbladian birth & \textit{BIRTH [L*]} & 4\\
Conjugate Lindbladian death & \textit{DEATH [L*]} & 4\\
Conjugate Lindbladian Swap & \textit{SWAP [L*]} & 4 \\
\hline
\end{tabular}
\end{center}
\caption {Values for meta-parameters describing the probabilities for the the different algorithm moves, used in the application to learning models for single and coupled quantum emitters, as discussed in the main text.}
\label{table:metaparameters_moves}
\end{table}

Additionally, we include meta parameters that govern the probabilities for the algorithm to pick any of the moves. The values, reported in Table \ref{table:metaparameters_moves}, were chosen by some initial optimization of the algorithm performance, using 47 independent runs, coarsely varying the move's proposal probabilities. When then used an additional set of 32 runs for finer optimization. During this procedure we considered several factors: mixing, mean accuracy, length of burn-in, number of processes per model, number of clusters, etc. During the finer-optimization step, diminishing returns were seen as evidence that an optimum had been reached.

\section{Mixing of Markov chains}
One of the key performance indicators of our algorithm, based on the construction of Markov-chains, is that chains adequately explore the model space, decoupling themselves from the point of initialization or from local optima of the likelihood. A possible way to test this is to utilize multiple chains, initialized in different random points, and assess how they explore the parameter space. Provided that the chains are long enough, they should converge to similar regions where the models explain well the experimental data. We define the overlap between chains as \textit{``mixing''} of the chains.

This task requires finding a balance between the exploration of new models (i.e. adding/removing operators, etc) and the learning of the parameter values for an existing model. With insufficient exploration, the models produced by the learning will not be representative of the space. On the contrary, without sufficient optimization of the parameter learning step for a given model, the model may be found unworthy (and discarded) based on a set of rates far from optimal.

The mixing of the chains can be affected by multiple meta-parameters, in particular the probabilities for the proposing rate updates compared to the other ``operator update'' moves. 

\begin{figure}[h]
    \centering
    \includegraphics[width=0.45\textwidth]{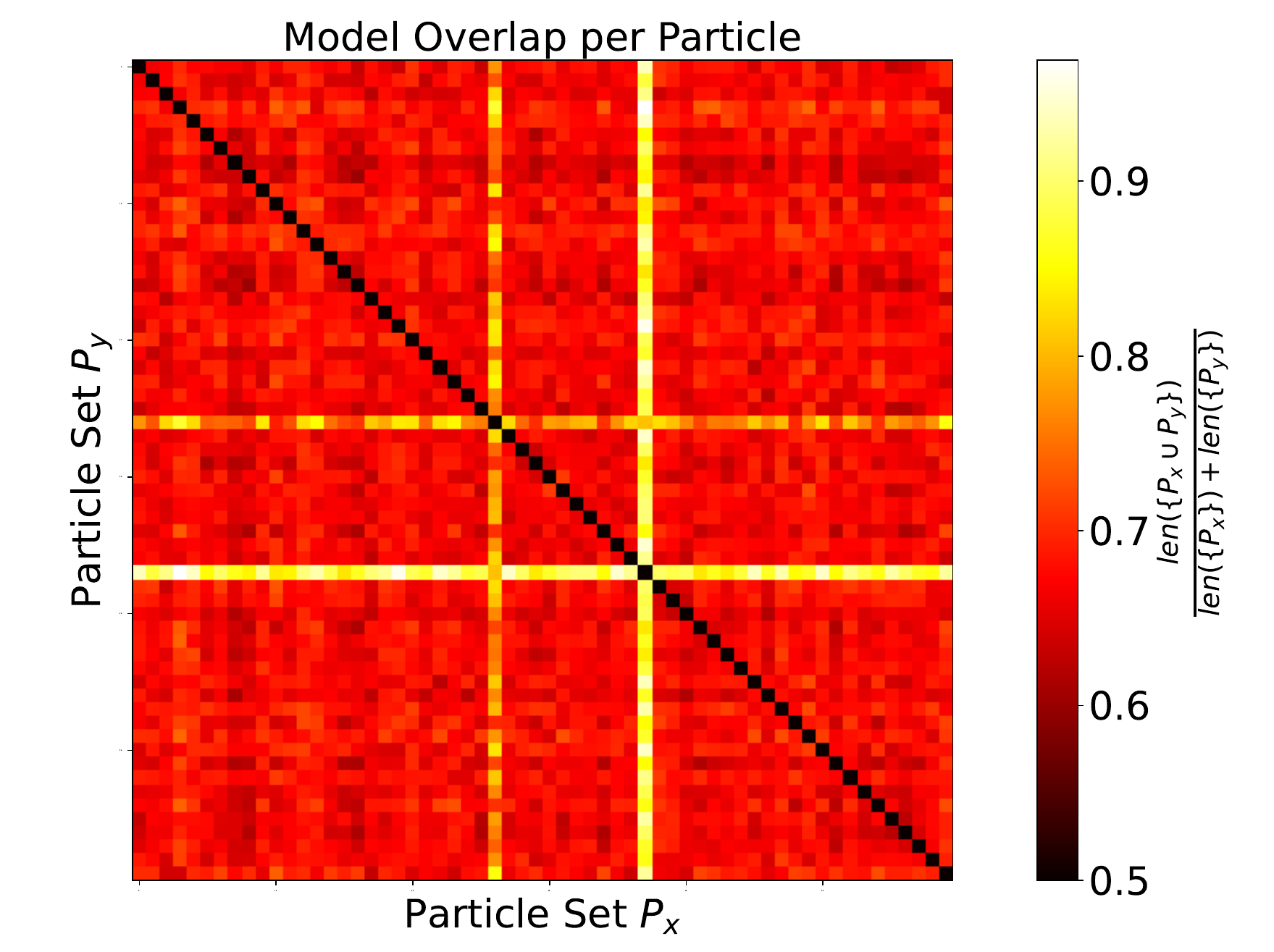}
    \caption{\textbf{Example of evaluation of mixing of Markov Chains.} Plot of $\mu (P_x, P_y)$, defined in Eq. ~\ref{eq:mixing}, for a set of chains, based on 60 independent particles. The example here showcases a high level of mixing between all particles, important to ensure that the parameter space is suitable explored. Features of note include two stand-alone particles with very little mixing with other chains and an average of 69\% between all chains.}
    \label{fig:mixing figure}
\end{figure}

In order to visualize mixing between different chains (or particles), we generate a number of parallel chains, we consider all pairs of particles $\lbrace P_x, P_y \rbrace$, and compute the quantity:
\begin{equation}
\label{eq:mixing}
    \mu (P_x, P_y) = \frac{len(\{P_x \cup P_y\})}{len(\{P_x\})+len(\{P_y\})}, 
\end{equation}
where $len(\cdot)$ denotes the number of models (we do not consider the corresponding rates here for simplicity) explored by the chain after burn-in. 
Each particle chosen as $P_x$ (or $P_y$) includes a set of model operators. If the two models chosen for $P_x$ and $P_y$ include completely different operators, with no overlap, then $len (P_x \cup P_y) = len (P_x) + len(P_y)$ and $\mu = 1$. In the other extreme case where $P_x$ and $P_y$ explore the same models, then $len (P_x \cup P_y) = len (P_x) = len(P_y)$ and $\mu = 1/2$. 

In Fig. (\ref{fig:mixing figure}), we plot $\mu$ for a well-mixed set of learning chains, based on 60 independent particles. Along the diagonal, $\mu = 0.5$ as there is complete overlap between a model and itself. Generally, a large variation in $\mu$ may suggest that the burn-in had not yet completed and that a chain is stuck. In Fig. (\ref{fig:mixing figure}), two chains stand out as having extremely low overlap with the rest, and thus not mixing well: these appear as statistical anomalies and might be safely removed.

\clearpage 


\begin{thebibliography}{82}%
\makeatletter
\providecommand \@ifxundefined [1]{%
 \@ifx{#1\undefined}
}%
\providecommand \@ifnum [1]{%
 \ifnum #1\expandafter \@firstoftwo
 \else \expandafter \@secondoftwo
 \fi
}%
\providecommand \@ifx [1]{%
 \ifx #1\expandafter \@firstoftwo
 \else \expandafter \@secondoftwo
 \fi
}%
\providecommand \natexlab [1]{#1}%
\providecommand \enquote  [1]{``#1''}%
\providecommand \bibnamefont  [1]{#1}%
\providecommand \bibfnamefont [1]{#1}%
\providecommand \citenamefont [1]{#1}%
\providecommand \href@noop [0]{\@secondoftwo}%
\providecommand \href [0]{\begingroup \@sanitize@url \@href}%
\providecommand \@href[1]{\@@startlink{#1}\@@href}%
\providecommand \@@href[1]{\endgroup#1\@@endlink}%
\providecommand \@sanitize@url [0]{\catcode `\\12\catcode `\$12\catcode `\&12\catcode `\#12\catcode `\^12\catcode `\_12\catcode `\%12\relax}%
\providecommand \@@startlink[1]{}%
\providecommand \@@endlink[0]{}%
\providecommand \url  [0]{\begingroup\@sanitize@url \@url }%
\providecommand \@url [1]{\endgroup\@href {#1}{\urlprefix }}%
\providecommand \urlprefix  [0]{URL }%
\providecommand \Eprint [0]{\href }%
\providecommand \doibase [0]{https://doi.org/}%
\providecommand \selectlanguage [0]{\@gobble}%
\providecommand \bibinfo  [0]{\@secondoftwo}%
\providecommand \bibfield  [0]{\@secondoftwo}%
\providecommand \translation [1]{[#1]}%
\providecommand \BibitemOpen [0]{}%
\providecommand \bibitemStop [0]{}%
\providecommand \bibitemNoStop [0]{.\EOS\space}%
\providecommand \EOS [0]{\spacefactor3000\relax}%
\providecommand \BibitemShut  [1]{\csname bibitem#1\endcsname}%
\let\auto@bib@innerbib\@empty
\bibitem [{\citenamefont {Butler}\ \emph {et~al.}(2018)\citenamefont {Butler}, \citenamefont {Davies}, \citenamefont {Cartwright}, \citenamefont {Isayev},\ and\ \citenamefont {Walsh}}]{butler_machine_2018}%
  \BibitemOpen
  \bibfield  {author} {\bibinfo {author} {\bibfnamefont {K.~T.}\ \bibnamefont {Butler}}, \bibinfo {author} {\bibfnamefont {D.~W.}\ \bibnamefont {Davies}}, \bibinfo {author} {\bibfnamefont {H.}~\bibnamefont {Cartwright}}, \bibinfo {author} {\bibfnamefont {O.}~\bibnamefont {Isayev}},\ and\ \bibinfo {author} {\bibfnamefont {A.}~\bibnamefont {Walsh}},\ }\href {https://doi.org/10.1038/s41586-018-0337-2} {\bibfield  {journal} {\bibinfo  {journal} {Nature}\ }\textbf {\bibinfo {volume} {559}},\ \bibinfo {pages} {547} (\bibinfo {year} {2018})}\BibitemShut {NoStop}%
\bibitem [{\citenamefont {Carleo}\ \emph {et~al.}(2019)\citenamefont {Carleo}, \citenamefont {Cirac}, \citenamefont {Cranmer}, \citenamefont {Daudet}, \citenamefont {Schuld}, \citenamefont {Tishby}, \citenamefont {Vogt-Maranto},\ and\ \citenamefont {Zdeborová}}]{carleo_machine_2019}%
  \BibitemOpen
  \bibfield  {author} {\bibinfo {author} {\bibfnamefont {G.}~\bibnamefont {Carleo}}, \bibinfo {author} {\bibfnamefont {I.}~\bibnamefont {Cirac}}, \bibinfo {author} {\bibfnamefont {K.}~\bibnamefont {Cranmer}}, \bibinfo {author} {\bibfnamefont {L.}~\bibnamefont {Daudet}}, \bibinfo {author} {\bibfnamefont {M.}~\bibnamefont {Schuld}}, \bibinfo {author} {\bibfnamefont {N.}~\bibnamefont {Tishby}}, \bibinfo {author} {\bibfnamefont {L.}~\bibnamefont {Vogt-Maranto}},\ and\ \bibinfo {author} {\bibfnamefont {L.}~\bibnamefont {Zdeborová}},\ }\href {https://doi.org/10.1103/RevModPhys.91.045002} {\bibfield  {journal} {\bibinfo  {journal} {Reviews of Modern Physics}\ }\textbf {\bibinfo {volume} {91}},\ \bibinfo {pages} {045002} (\bibinfo {year} {2019})}\BibitemShut {NoStop}%
\bibitem [{\citenamefont {Moosavi}\ \emph {et~al.}(2020)\citenamefont {Moosavi}, \citenamefont {Jablonka},\ and\ \citenamefont {Smit}}]{moosavi_role_2020}%
  \BibitemOpen
  \bibfield  {author} {\bibinfo {author} {\bibfnamefont {S.~M.}\ \bibnamefont {Moosavi}}, \bibinfo {author} {\bibfnamefont {K.~M.}\ \bibnamefont {Jablonka}},\ and\ \bibinfo {author} {\bibfnamefont {B.}~\bibnamefont {Smit}},\ }\href {https://doi.org/10.1021/jacs.0c09105} {\bibfield  {journal} {\bibinfo  {journal} {Journal of the American Chemical Society}\ }\textbf {\bibinfo {volume} {142}},\ \bibinfo {pages} {20273} (\bibinfo {year} {2020})}\BibitemShut {NoStop}%
\bibitem [{\citenamefont {Krenn}\ \emph {et~al.}(2022)\citenamefont {Krenn}, \citenamefont {Pollice}, \citenamefont {Guo}, \citenamefont {Aldeghi}, \citenamefont {Cervera-Lierta}, \citenamefont {Friederich}, \citenamefont {dos Passos Gomes}, \citenamefont {Häse}, \citenamefont {Jinich}, \citenamefont {Nigam}, \citenamefont {Yao},\ and\ \citenamefont {Aspuru-Guzik}}]{krenn_scientific_2022}%
  \BibitemOpen
  \bibfield  {author} {\bibinfo {author} {\bibfnamefont {M.}~\bibnamefont {Krenn}}, \bibinfo {author} {\bibfnamefont {R.}~\bibnamefont {Pollice}}, \bibinfo {author} {\bibfnamefont {S.~Y.}\ \bibnamefont {Guo}}, \bibinfo {author} {\bibfnamefont {M.}~\bibnamefont {Aldeghi}}, \bibinfo {author} {\bibfnamefont {A.}~\bibnamefont {Cervera-Lierta}}, \bibinfo {author} {\bibfnamefont {P.}~\bibnamefont {Friederich}}, \bibinfo {author} {\bibfnamefont {G.}~\bibnamefont {dos Passos Gomes}}, \bibinfo {author} {\bibfnamefont {F.}~\bibnamefont {Häse}}, \bibinfo {author} {\bibfnamefont {A.}~\bibnamefont {Jinich}}, \bibinfo {author} {\bibfnamefont {A.}~\bibnamefont {Nigam}}, \bibinfo {author} {\bibfnamefont {Z.}~\bibnamefont {Yao}},\ and\ \bibinfo {author} {\bibfnamefont {A.}~\bibnamefont {Aspuru-Guzik}},\ }\href {https://doi.org/10.1038/s42254-022-00518-3} {\bibfield  {journal} {\bibinfo  {journal} {Nature Reviews Physics}\ }\textbf {\bibinfo {volume} {4}},\ \bibinfo {pages} {761} (\bibinfo {year} {2022})}\BibitemShut
  {NoStop}%
\bibitem [{\citenamefont {Gebhart}\ \emph {et~al.}(2023)\citenamefont {Gebhart}, \citenamefont {Santagati}, \citenamefont {Gentile}, \citenamefont {Gauger}, \citenamefont {Craig}, \citenamefont {Ares}, \citenamefont {Banchi}, \citenamefont {Marquardt}, \citenamefont {Pezzè},\ and\ \citenamefont {Bonato}}]{gebhart_learning_2023}%
  \BibitemOpen
  \bibfield  {author} {\bibinfo {author} {\bibfnamefont {V.}~\bibnamefont {Gebhart}}, \bibinfo {author} {\bibfnamefont {R.}~\bibnamefont {Santagati}}, \bibinfo {author} {\bibfnamefont {A.~A.}\ \bibnamefont {Gentile}}, \bibinfo {author} {\bibfnamefont {E.~M.}\ \bibnamefont {Gauger}}, \bibinfo {author} {\bibfnamefont {D.}~\bibnamefont {Craig}}, \bibinfo {author} {\bibfnamefont {N.}~\bibnamefont {Ares}}, \bibinfo {author} {\bibfnamefont {L.}~\bibnamefont {Banchi}}, \bibinfo {author} {\bibfnamefont {F.}~\bibnamefont {Marquardt}}, \bibinfo {author} {\bibfnamefont {L.}~\bibnamefont {Pezzè}},\ and\ \bibinfo {author} {\bibfnamefont {C.}~\bibnamefont {Bonato}},\ }\href {https://doi.org/10.1038/s42254-022-00552-1} {\bibfield  {journal} {\bibinfo  {journal} {Nature Reviews Physics}\ }\textbf {\bibinfo {volume} {5}},\ \bibinfo {pages} {141} (\bibinfo {year} {2023})}\BibitemShut {NoStop}%
\bibitem [{\citenamefont {Krenn}\ \emph {et~al.}(2023)\citenamefont {Krenn}, \citenamefont {Landgraf}, \citenamefont {Foesel},\ and\ \citenamefont {Marquardt}}]{krenn_artificial_2023}%
  \BibitemOpen
  \bibfield  {author} {\bibinfo {author} {\bibfnamefont {M.}~\bibnamefont {Krenn}}, \bibinfo {author} {\bibfnamefont {J.}~\bibnamefont {Landgraf}}, \bibinfo {author} {\bibfnamefont {T.}~\bibnamefont {Foesel}},\ and\ \bibinfo {author} {\bibfnamefont {F.}~\bibnamefont {Marquardt}},\ }\href {https://doi.org/10.1103/PhysRevA.107.010101} {\bibfield  {journal} {\bibinfo  {journal} {Physical Review A}\ }\textbf {\bibinfo {volume} {107}},\ \bibinfo {pages} {010101} (\bibinfo {year} {2023})}\BibitemShut {NoStop}%
\bibitem [{\citenamefont {Usman}\ \emph {et~al.}(2020)\citenamefont {Usman}, \citenamefont {Wong}, \citenamefont {Hill},\ and\ \citenamefont {Hollenberg}}]{usman_framework_2020}%
  \BibitemOpen
  \bibfield  {author} {\bibinfo {author} {\bibfnamefont {M.}~\bibnamefont {Usman}}, \bibinfo {author} {\bibfnamefont {Y.~Z.}\ \bibnamefont {Wong}}, \bibinfo {author} {\bibfnamefont {C.~D.}\ \bibnamefont {Hill}},\ and\ \bibinfo {author} {\bibfnamefont {L.~C.~L.}\ \bibnamefont {Hollenberg}},\ }\href {https://doi.org/10.1038/s41524-020-0282-0} {\bibfield  {journal} {\bibinfo  {journal} {npj Computational Materials}\ }\textbf {\bibinfo {volume} {6}},\ \bibinfo {pages} {1} (\bibinfo {year} {2020})}\BibitemShut {NoStop}%
\bibitem [{\citenamefont {Moon}\ \emph {et~al.}(2020)\citenamefont {Moon}, \citenamefont {Lennon}, \citenamefont {Kirkpatrick}, \citenamefont {van Esbroeck}, \citenamefont {Camenzind}, \citenamefont {Yu}, \citenamefont {Vigneau}, \citenamefont {Zumbühl}, \citenamefont {Briggs}, \citenamefont {Osborne}, \citenamefont {Sejdinovic}, \citenamefont {Laird},\ and\ \citenamefont {Ares}}]{moon_machine_2020}%
  \BibitemOpen
  \bibfield  {author} {\bibinfo {author} {\bibfnamefont {H.}~\bibnamefont {Moon}}, \bibinfo {author} {\bibfnamefont {D.~T.}\ \bibnamefont {Lennon}}, \bibinfo {author} {\bibfnamefont {J.}~\bibnamefont {Kirkpatrick}}, \bibinfo {author} {\bibfnamefont {N.~M.}\ \bibnamefont {van Esbroeck}}, \bibinfo {author} {\bibfnamefont {L.~C.}\ \bibnamefont {Camenzind}}, \bibinfo {author} {\bibfnamefont {L.}~\bibnamefont {Yu}}, \bibinfo {author} {\bibfnamefont {F.}~\bibnamefont {Vigneau}}, \bibinfo {author} {\bibfnamefont {D.~M.}\ \bibnamefont {Zumbühl}}, \bibinfo {author} {\bibfnamefont {G.~a.~D.}\ \bibnamefont {Briggs}}, \bibinfo {author} {\bibfnamefont {M.~A.}\ \bibnamefont {Osborne}}, \bibinfo {author} {\bibfnamefont {D.}~\bibnamefont {Sejdinovic}}, \bibinfo {author} {\bibfnamefont {E.~A.}\ \bibnamefont {Laird}},\ and\ \bibinfo {author} {\bibfnamefont {N.}~\bibnamefont {Ares}},\ }\href {https://doi.org/10.1038/s41467-020-17835-9} {\bibfield  {journal} {\bibinfo  {journal} {Nature Communications}\ }\textbf {\bibinfo
  {volume} {11}},\ \bibinfo {pages} {4161} (\bibinfo {year} {2020})}\BibitemShut {NoStop}%
\bibitem [{\citenamefont {Lee}\ \emph {et~al.}(2021)\citenamefont {Lee}, \citenamefont {Nam}, \citenamefont {Ji}, \citenamefont {Choi}, \citenamefont {Jin}, \citenamefont {Kim}, \citenamefont {Na}, \citenamefont {Ryu}, \citenamefont {Cho}, \citenamefont {Lee}, \citenamefont {Lee}, \citenamefont {Joo},\ and\ \citenamefont {Kim}}]{lee_multiple_2021}%
  \BibitemOpen
  \bibfield  {author} {\bibinfo {author} {\bibfnamefont {K.}~\bibnamefont {Lee}}, \bibinfo {author} {\bibfnamefont {S.}~\bibnamefont {Nam}}, \bibinfo {author} {\bibfnamefont {H.}~\bibnamefont {Ji}}, \bibinfo {author} {\bibfnamefont {J.}~\bibnamefont {Choi}}, \bibinfo {author} {\bibfnamefont {J.-E.}\ \bibnamefont {Jin}}, \bibinfo {author} {\bibfnamefont {Y.}~\bibnamefont {Kim}}, \bibinfo {author} {\bibfnamefont {J.}~\bibnamefont {Na}}, \bibinfo {author} {\bibfnamefont {M.-Y.}\ \bibnamefont {Ryu}}, \bibinfo {author} {\bibfnamefont {Y.-H.}\ \bibnamefont {Cho}}, \bibinfo {author} {\bibfnamefont {H.}~\bibnamefont {Lee}}, \bibinfo {author} {\bibfnamefont {J.}~\bibnamefont {Lee}}, \bibinfo {author} {\bibfnamefont {M.-K.}\ \bibnamefont {Joo}},\ and\ \bibinfo {author} {\bibfnamefont {G.-T.}\ \bibnamefont {Kim}},\ }\href {https://doi.org/10.1038/s41699-020-00186-w} {\bibfield  {journal} {\bibinfo  {journal} {npj 2D Materials and Applications}\ }\textbf {\bibinfo {volume} {5}},\ \bibinfo {pages} {1} (\bibinfo {year}
  {2021})}\BibitemShut {NoStop}%
\bibitem [{\citenamefont {Valenti}\ \emph {et~al.}(2022)\citenamefont {Valenti}, \citenamefont {Jin}, \citenamefont {Léonard}, \citenamefont {Huber},\ and\ \citenamefont {Greplova}}]{valenti_scalable_2022}%
  \BibitemOpen
  \bibfield  {author} {\bibinfo {author} {\bibfnamefont {A.}~\bibnamefont {Valenti}}, \bibinfo {author} {\bibfnamefont {G.}~\bibnamefont {Jin}}, \bibinfo {author} {\bibfnamefont {J.}~\bibnamefont {Léonard}}, \bibinfo {author} {\bibfnamefont {S.~D.}\ \bibnamefont {Huber}},\ and\ \bibinfo {author} {\bibfnamefont {E.}~\bibnamefont {Greplova}},\ }\href {https://doi.org/10.1103/PhysRevA.105.023302} {\bibfield  {journal} {\bibinfo  {journal} {Physical Review A}\ }\textbf {\bibinfo {volume} {105}},\ \bibinfo {pages} {023302} (\bibinfo {year} {2022})}\BibitemShut {NoStop}%
\bibitem [{\citenamefont {Thomas}\ \emph {et~al.}(2023)\citenamefont {Thomas}, \citenamefont {Ciriano-Tejel}, \citenamefont {Wise}, \citenamefont {Prete}, \citenamefont {de~Kruijf}, \citenamefont {Ibberson}, \citenamefont {Noah}, \citenamefont {Gomez-Saiz}, \citenamefont {Gonzalez-Zalba}, \citenamefont {Johnson},\ and\ \citenamefont {Morton}}]{thomas_rapid_2023}%
  \BibitemOpen
  \bibfield  {author} {\bibinfo {author} {\bibfnamefont {E.~J.}\ \bibnamefont {Thomas}}, \bibinfo {author} {\bibfnamefont {V.~N.}\ \bibnamefont {Ciriano-Tejel}}, \bibinfo {author} {\bibfnamefont {D.~F.}\ \bibnamefont {Wise}}, \bibinfo {author} {\bibfnamefont {D.}~\bibnamefont {Prete}}, \bibinfo {author} {\bibfnamefont {M.}~\bibnamefont {de~Kruijf}}, \bibinfo {author} {\bibfnamefont {D.~J.}\ \bibnamefont {Ibberson}}, \bibinfo {author} {\bibfnamefont {G.~M.}\ \bibnamefont {Noah}}, \bibinfo {author} {\bibfnamefont {A.}~\bibnamefont {Gomez-Saiz}}, \bibinfo {author} {\bibfnamefont {M.~F.}\ \bibnamefont {Gonzalez-Zalba}}, \bibinfo {author} {\bibfnamefont {M.~A.~I.}\ \bibnamefont {Johnson}},\ and\ \bibinfo {author} {\bibfnamefont {J.~J.~L.}\ \bibnamefont {Morton}},\ }\href {https://doi.org/10.48550/arXiv.2310.20434} {\bibinfo {title} {Rapid cryogenic characterisation of 1024 integrated silicon quantum dots}} (\bibinfo {year} {2023})\BibitemShut {NoStop}%
\bibitem [{\citenamefont {Craig}\ \emph {et~al.}(2024{\natexlab{a}})\citenamefont {Craig}, \citenamefont {Moon}, \citenamefont {Fedele}, \citenamefont {Lennon}, \citenamefont {van Straaten}, \citenamefont {Vigneau}, \citenamefont {Camenzind}, \citenamefont {Zumbühl}, \citenamefont {Briggs}, \citenamefont {Osborne}, \citenamefont {Sejdinovic},\ and\ \citenamefont {Ares}}]{craig_bridging_2024}%
  \BibitemOpen
  \bibfield  {author} {\bibinfo {author} {\bibfnamefont {D.}~\bibnamefont {Craig}}, \bibinfo {author} {\bibfnamefont {H.}~\bibnamefont {Moon}}, \bibinfo {author} {\bibfnamefont {F.}~\bibnamefont {Fedele}}, \bibinfo {author} {\bibfnamefont {D.}~\bibnamefont {Lennon}}, \bibinfo {author} {\bibfnamefont {B.}~\bibnamefont {van Straaten}}, \bibinfo {author} {\bibfnamefont {F.}~\bibnamefont {Vigneau}}, \bibinfo {author} {\bibfnamefont {L.}~\bibnamefont {Camenzind}}, \bibinfo {author} {\bibfnamefont {D.}~\bibnamefont {Zumbühl}}, \bibinfo {author} {\bibfnamefont {G.}~\bibnamefont {Briggs}}, \bibinfo {author} {\bibfnamefont {M.}~\bibnamefont {Osborne}}, \bibinfo {author} {\bibfnamefont {D.}~\bibnamefont {Sejdinovic}},\ and\ \bibinfo {author} {\bibfnamefont {N.}~\bibnamefont {Ares}},\ }\href {https://doi.org/10.1103/PhysRevX.14.011001} {\bibfield  {journal} {\bibinfo  {journal} {Physical Review X}\ }\textbf {\bibinfo {volume} {14}},\ \bibinfo {pages} {011001} (\bibinfo {year} {2024}{\natexlab{a}})}\BibitemShut
  {NoStop}%
\bibitem [{\citenamefont {Schmidt}\ and\ \citenamefont {Lipson}(2009)}]{schmidt_distilling_2009}%
  \BibitemOpen
  \bibfield  {author} {\bibinfo {author} {\bibfnamefont {M.}~\bibnamefont {Schmidt}}\ and\ \bibinfo {author} {\bibfnamefont {H.}~\bibnamefont {Lipson}},\ }\href {https://doi.org/10.1126/science.1165893} {\bibfield  {journal} {\bibinfo  {journal} {Science}\ }\textbf {\bibinfo {volume} {324}},\ \bibinfo {pages} {81} (\bibinfo {year} {2009})}\BibitemShut {NoStop}%
\bibitem [{\citenamefont {Liu}\ and\ \citenamefont {Tegmark}(2021)}]{liu_machine_2021}%
  \BibitemOpen
  \bibfield  {author} {\bibinfo {author} {\bibfnamefont {Z.}~\bibnamefont {Liu}}\ and\ \bibinfo {author} {\bibfnamefont {M.}~\bibnamefont {Tegmark}},\ }\bibfield  {journal} {\bibinfo  {journal} {Physical Review Letters}\ }\href {https://doi.org/10.1103/physrevlett.126.180604} {10.1103/physrevlett.126.180604} (\bibinfo {year} {2021})\BibitemShut {NoStop}%
\bibitem [{\citenamefont {Lemos}\ \emph {et~al.}(2023)\citenamefont {Lemos}, \citenamefont {Jeffrey}, \citenamefont {Cranmer}, \citenamefont {Ho},\ and\ \citenamefont {Battaglia}}]{lemos_rediscovering_2023}%
  \BibitemOpen
  \bibfield  {author} {\bibinfo {author} {\bibfnamefont {P.}~\bibnamefont {Lemos}}, \bibinfo {author} {\bibfnamefont {N.}~\bibnamefont {Jeffrey}}, \bibinfo {author} {\bibfnamefont {M.}~\bibnamefont {Cranmer}}, \bibinfo {author} {\bibfnamefont {S.}~\bibnamefont {Ho}},\ and\ \bibinfo {author} {\bibfnamefont {P.}~\bibnamefont {Battaglia}},\ }\href {https://doi.org/10.1088/2632-2153/acfa63} {\bibfield  {journal} {\bibinfo  {journal} {Machine Learning: Science and Technology}\ }\textbf {\bibinfo {volume} {4}},\ \bibinfo {pages} {045002} (\bibinfo {year} {2023})}\BibitemShut {NoStop}%
\bibitem [{\citenamefont {Chen}\ \emph {et~al.}(2024)\citenamefont {Chen}, \citenamefont {Soh}, \citenamefont {Ooi}, \citenamefont {Vissol-Gaudin}, \citenamefont {Yu}, \citenamefont {Novoselov}, \citenamefont {Hippalgaonkar},\ and\ \citenamefont {Li}}]{chen_constructing_2024}%
  \BibitemOpen
  \bibfield  {author} {\bibinfo {author} {\bibfnamefont {X.}~\bibnamefont {Chen}}, \bibinfo {author} {\bibfnamefont {B.~W.}\ \bibnamefont {Soh}}, \bibinfo {author} {\bibfnamefont {Z.-E.}\ \bibnamefont {Ooi}}, \bibinfo {author} {\bibfnamefont {E.}~\bibnamefont {Vissol-Gaudin}}, \bibinfo {author} {\bibfnamefont {H.}~\bibnamefont {Yu}}, \bibinfo {author} {\bibfnamefont {K.~S.}\ \bibnamefont {Novoselov}}, \bibinfo {author} {\bibfnamefont {K.}~\bibnamefont {Hippalgaonkar}},\ and\ \bibinfo {author} {\bibfnamefont {Q.}~\bibnamefont {Li}},\ }\href {https://doi.org/10.1038/s43588-023-00581-5} {\bibfield  {journal} {\bibinfo  {journal} {Nature Computational Science}\ }\textbf {\bibinfo {volume} {4}},\ \bibinfo {pages} {66} (\bibinfo {year} {2024})}\BibitemShut {NoStop}%
\bibitem [{\citenamefont {Torlai}\ and\ \citenamefont {Melko}(2016)}]{torlai_learning_2016}%
  \BibitemOpen
  \bibfield  {author} {\bibinfo {author} {\bibfnamefont {G.}~\bibnamefont {Torlai}}\ and\ \bibinfo {author} {\bibfnamefont {R.~G.}\ \bibnamefont {Melko}},\ }\href {https://doi.org/10.1103/PhysRevB.94.165134} {\bibfield  {journal} {\bibinfo  {journal} {Physical Review B}\ }\textbf {\bibinfo {volume} {94}},\ \bibinfo {pages} {165134} (\bibinfo {year} {2016})}\BibitemShut {NoStop}%
\bibitem [{\citenamefont {van Nieuwenburg}\ \emph {et~al.}(2017)\citenamefont {van Nieuwenburg}, \citenamefont {Liu},\ and\ \citenamefont {Huber}}]{van_nieuwenburg_learning_2017}%
  \BibitemOpen
  \bibfield  {author} {\bibinfo {author} {\bibfnamefont {E.~P.~L.}\ \bibnamefont {van Nieuwenburg}}, \bibinfo {author} {\bibfnamefont {Y.-H.}\ \bibnamefont {Liu}},\ and\ \bibinfo {author} {\bibfnamefont {S.~D.}\ \bibnamefont {Huber}},\ }\href {https://doi.org/10.1038/nphys4037} {\bibfield  {journal} {\bibinfo  {journal} {Nature Physics}\ }\textbf {\bibinfo {volume} {13}},\ \bibinfo {pages} {435} (\bibinfo {year} {2017})}\BibitemShut {NoStop}%
\bibitem [{\citenamefont {Carrasquilla}\ \emph {et~al.}(2019)\citenamefont {Carrasquilla}, \citenamefont {Torlai}, \citenamefont {Melko},\ and\ \citenamefont {Aolita}}]{carrasquilla_reconstructing_2019}%
  \BibitemOpen
  \bibfield  {author} {\bibinfo {author} {\bibfnamefont {J.}~\bibnamefont {Carrasquilla}}, \bibinfo {author} {\bibfnamefont {G.}~\bibnamefont {Torlai}}, \bibinfo {author} {\bibfnamefont {R.~G.}\ \bibnamefont {Melko}},\ and\ \bibinfo {author} {\bibfnamefont {L.}~\bibnamefont {Aolita}},\ }\href {https://doi.org/10.1038/s42256-019-0028-1} {\bibfield  {journal} {\bibinfo  {journal} {Nature Machine Intelligence}\ }\textbf {\bibinfo {volume} {1}},\ \bibinfo {pages} {155} (\bibinfo {year} {2019})}\BibitemShut {NoStop}%
\bibitem [{\citenamefont {Torlai}\ \emph {et~al.}(2019)\citenamefont {Torlai}, \citenamefont {Timar}, \citenamefont {van Nieuwenburg}, \citenamefont {Levine}, \citenamefont {Omran}, \citenamefont {Keesling}, \citenamefont {Bernien}, \citenamefont {Greiner}, \citenamefont {Vuletić}, \citenamefont {Lukin}, \citenamefont {Melko},\ and\ \citenamefont {Endres}}]{torlai_integrating_2019}%
  \BibitemOpen
  \bibfield  {author} {\bibinfo {author} {\bibfnamefont {G.}~\bibnamefont {Torlai}}, \bibinfo {author} {\bibfnamefont {B.}~\bibnamefont {Timar}}, \bibinfo {author} {\bibfnamefont {E.~P.}\ \bibnamefont {van Nieuwenburg}}, \bibinfo {author} {\bibfnamefont {H.}~\bibnamefont {Levine}}, \bibinfo {author} {\bibfnamefont {A.}~\bibnamefont {Omran}}, \bibinfo {author} {\bibfnamefont {A.}~\bibnamefont {Keesling}}, \bibinfo {author} {\bibfnamefont {H.}~\bibnamefont {Bernien}}, \bibinfo {author} {\bibfnamefont {M.}~\bibnamefont {Greiner}}, \bibinfo {author} {\bibfnamefont {V.}~\bibnamefont {Vuletić}}, \bibinfo {author} {\bibfnamefont {M.~D.}\ \bibnamefont {Lukin}}, \bibinfo {author} {\bibfnamefont {R.~G.}\ \bibnamefont {Melko}},\ and\ \bibinfo {author} {\bibfnamefont {M.}~\bibnamefont {Endres}},\ }\href {https://doi.org/10.1103/PhysRevLett.123.230504} {\bibfield  {journal} {\bibinfo  {journal} {Physical Review Letters}\ }\textbf {\bibinfo {volume} {123}},\ \bibinfo {pages} {230504} (\bibinfo {year} {2019})}\BibitemShut
  {NoStop}%
\bibitem [{\citenamefont {Gentile}\ \emph {et~al.}(2021)\citenamefont {Gentile}, \citenamefont {Flynn}, \citenamefont {Knauer}, \citenamefont {Wiebe}, \citenamefont {Paesani}, \citenamefont {Granade}, \citenamefont {Rarity}, \citenamefont {Santagati},\ and\ \citenamefont {Laing}}]{gentile_learning_2021}%
  \BibitemOpen
  \bibfield  {author} {\bibinfo {author} {\bibfnamefont {A.~A.}\ \bibnamefont {Gentile}}, \bibinfo {author} {\bibfnamefont {B.}~\bibnamefont {Flynn}}, \bibinfo {author} {\bibfnamefont {S.}~\bibnamefont {Knauer}}, \bibinfo {author} {\bibfnamefont {N.}~\bibnamefont {Wiebe}}, \bibinfo {author} {\bibfnamefont {S.}~\bibnamefont {Paesani}}, \bibinfo {author} {\bibfnamefont {C.~E.}\ \bibnamefont {Granade}}, \bibinfo {author} {\bibfnamefont {J.~G.}\ \bibnamefont {Rarity}}, \bibinfo {author} {\bibfnamefont {R.}~\bibnamefont {Santagati}},\ and\ \bibinfo {author} {\bibfnamefont {A.}~\bibnamefont {Laing}},\ }\href {https://doi.org/10.1038/s41567-021-01201-7} {\bibfield  {journal} {\bibinfo  {journal} {Nature Physics}\ ,\ \bibinfo {pages} {1}} (\bibinfo {year} {2021})}\BibitemShut {NoStop}%
\bibitem [{\citenamefont {Anshu}\ \emph {et~al.}(2021)\citenamefont {Anshu}, \citenamefont {Arunachalam}, \citenamefont {Kuwahara},\ and\ \citenamefont {Soleimanifar}}]{anshu_sample-efficient_2021}%
  \BibitemOpen
  \bibfield  {author} {\bibinfo {author} {\bibfnamefont {A.}~\bibnamefont {Anshu}}, \bibinfo {author} {\bibfnamefont {S.}~\bibnamefont {Arunachalam}}, \bibinfo {author} {\bibfnamefont {T.}~\bibnamefont {Kuwahara}},\ and\ \bibinfo {author} {\bibfnamefont {M.}~\bibnamefont {Soleimanifar}},\ }\href {https://doi.org/10.1038/s41567-021-01232-0} {\bibfield  {journal} {\bibinfo  {journal} {Nature Physics}\ ,\ \bibinfo {pages} {1}} (\bibinfo {year} {2021})}\BibitemShut {NoStop}%
\bibitem [{\citenamefont {Flam-Shepherd}\ \emph {et~al.}(2022)\citenamefont {Flam-Shepherd}, \citenamefont {Wu}, \citenamefont {Gu}, \citenamefont {Cervera-Lierta}, \citenamefont {Krenn},\ and\ \citenamefont {Aspuru-Guzik}}]{flam-shepherd_learning_2022}%
  \BibitemOpen
  \bibfield  {author} {\bibinfo {author} {\bibfnamefont {D.}~\bibnamefont {Flam-Shepherd}}, \bibinfo {author} {\bibfnamefont {T.~C.}\ \bibnamefont {Wu}}, \bibinfo {author} {\bibfnamefont {X.}~\bibnamefont {Gu}}, \bibinfo {author} {\bibfnamefont {A.}~\bibnamefont {Cervera-Lierta}}, \bibinfo {author} {\bibfnamefont {M.}~\bibnamefont {Krenn}},\ and\ \bibinfo {author} {\bibfnamefont {A.}~\bibnamefont {Aspuru-Guzik}},\ }\href {https://doi.org/10.1038/s42256-022-00493-5} {\bibfield  {journal} {\bibinfo  {journal} {Nature Machine Intelligence}\ }\textbf {\bibinfo {volume} {4}},\ \bibinfo {pages} {544} (\bibinfo {year} {2022})}\BibitemShut {NoStop}%
\bibitem [{\citenamefont {Arlt}\ \emph {et~al.}(2022)\citenamefont {Arlt}, \citenamefont {Ruiz-Gonzalez},\ and\ \citenamefont {Krenn}}]{arlt_digital_2022}%
  \BibitemOpen
  \bibfield  {author} {\bibinfo {author} {\bibfnamefont {S.}~\bibnamefont {Arlt}}, \bibinfo {author} {\bibfnamefont {C.}~\bibnamefont {Ruiz-Gonzalez}},\ and\ \bibinfo {author} {\bibfnamefont {M.}~\bibnamefont {Krenn}},\ }\href {https://doi.org/10.48550/arXiv.2210.09981} {\bibinfo {title} {Digital {Discovery} of a {Scientific} {Concept} at the {Core} of {Experimental} {Quantum} {Optics}}} (\bibinfo {year} {2022}),\ \bibinfo {note} {arXiv:2210.09981 [quant-ph]}\BibitemShut {NoStop}%
\bibitem [{\citenamefont {Huang}\ \emph {et~al.}(2022)\citenamefont {Huang}, \citenamefont {Kueng}, \citenamefont {Torlai}, \citenamefont {Albert},\ and\ \citenamefont {Preskill}}]{huang_provably_2022}%
  \BibitemOpen
  \bibfield  {author} {\bibinfo {author} {\bibfnamefont {H.-Y.}\ \bibnamefont {Huang}}, \bibinfo {author} {\bibfnamefont {R.}~\bibnamefont {Kueng}}, \bibinfo {author} {\bibfnamefont {G.}~\bibnamefont {Torlai}}, \bibinfo {author} {\bibfnamefont {V.~V.}\ \bibnamefont {Albert}},\ and\ \bibinfo {author} {\bibfnamefont {J.}~\bibnamefont {Preskill}},\ }\bibfield  {journal} {\bibinfo  {journal} {Science}\ }\textbf {\bibinfo {volume} {377}},\ \href {https://doi.org/10.1126/science.abk3333} {10.1126/science.abk3333} (\bibinfo {year} {2022})\BibitemShut {NoStop}%
\bibitem [{\citenamefont {Breuer}\ and\ \citenamefont {Petruccione}(2007)}]{breuer_theory_2007}%
  \BibitemOpen
  \bibfield  {author} {\bibinfo {author} {\bibfnamefont {H.-P.}\ \bibnamefont {Breuer}}\ and\ \bibinfo {author} {\bibfnamefont {F.}~\bibnamefont {Petruccione}},\ }\href {https://doi.org/10.1093/acprof:oso/9780199213900.001.0001} {\emph {\bibinfo {title} {The {Theory} of {Open} {Quantum} {Systems}}}},\ \bibinfo {edition} {1st}\ ed.\ (\bibinfo  {publisher} {Oxford University PressOxford},\ \bibinfo {year} {2007})\BibitemShut {NoStop}%
\bibitem [{\citenamefont {Lanyon}\ \emph {et~al.}(2017)\citenamefont {Lanyon}, \citenamefont {Maier}, \citenamefont {Holzäpfel}, \citenamefont {Baumgratz}, \citenamefont {Hempel}, \citenamefont {Jurcevic}, \citenamefont {Dhand}, \citenamefont {Buyskikh}, \citenamefont {Daley}, \citenamefont {Cramer}, \citenamefont {Plenio}, \citenamefont {Blatt}, ,\ and\ \citenamefont {Roos}}]{b_p_lanyon_efficient_2017}%
  \BibitemOpen
  \bibfield  {author} {\bibinfo {author} {\bibfnamefont {B.~P.}\ \bibnamefont {Lanyon}}, \bibinfo {author} {\bibfnamefont {C.}~\bibnamefont {Maier}}, \bibinfo {author} {\bibfnamefont {M.}~\bibnamefont {Holzäpfel}}, \bibinfo {author} {\bibfnamefont {T.}~\bibnamefont {Baumgratz}}, \bibinfo {author} {\bibfnamefont {C.}~\bibnamefont {Hempel}}, \bibinfo {author} {\bibfnamefont {P.}~\bibnamefont {Jurcevic}}, \bibinfo {author} {\bibfnamefont {I.}~\bibnamefont {Dhand}}, \bibinfo {author} {\bibfnamefont {A.~S.}\ \bibnamefont {Buyskikh}}, \bibinfo {author} {\bibfnamefont {A.~J.}\ \bibnamefont {Daley}}, \bibinfo {author} {\bibfnamefont {M.}~\bibnamefont {Cramer}}, \bibinfo {author} {\bibfnamefont {M.~B.}\ \bibnamefont {Plenio}}, \bibinfo {author} {\bibfnamefont {R.}~\bibnamefont {Blatt}}, ,\ and\ \bibinfo {author} {\bibfnamefont {C.}~\bibnamefont {Roos}},\ }\href {https://doi.org/10.1038/nphys4244} {\bibfield  {journal} {\bibinfo  {journal} {Nature Physics}\ }\textbf {\bibinfo {volume} {13}},\ \bibinfo {pages} {1158}
  (\bibinfo {year} {2017})}\BibitemShut {NoStop}%
\bibitem [{\citenamefont {Granade}\ \emph {et~al.}(2016)\citenamefont {Granade}, \citenamefont {Combes},\ and\ \citenamefont {Cory}}]{granade_practical_2016}%
  \BibitemOpen
  \bibfield  {author} {\bibinfo {author} {\bibfnamefont {C.}~\bibnamefont {Granade}}, \bibinfo {author} {\bibfnamefont {J.}~\bibnamefont {Combes}},\ and\ \bibinfo {author} {\bibfnamefont {D.~G.}\ \bibnamefont {Cory}},\ }\href {https://doi.org/10.1088/1367-2630/18/3/033024} {\bibfield  {journal} {\bibinfo  {journal} {New Journal of Physics}\ }\textbf {\bibinfo {volume} {18}},\ \bibinfo {pages} {033024} (\bibinfo {year} {2016})}\BibitemShut {NoStop}%
\bibitem [{\citenamefont {Pollock}\ \emph {et~al.}(2018{\natexlab{a}})\citenamefont {Pollock}, \citenamefont {Rodríguez-Rosario}, \citenamefont {Frauenheim}, \citenamefont {Paternostro},\ and\ \citenamefont {Modi}}]{pollock_non-markovian_2018}%
  \BibitemOpen
  \bibfield  {author} {\bibinfo {author} {\bibfnamefont {F.~A.}\ \bibnamefont {Pollock}}, \bibinfo {author} {\bibfnamefont {C.}~\bibnamefont {Rodríguez-Rosario}}, \bibinfo {author} {\bibfnamefont {T.}~\bibnamefont {Frauenheim}}, \bibinfo {author} {\bibfnamefont {M.}~\bibnamefont {Paternostro}},\ and\ \bibinfo {author} {\bibfnamefont {K.}~\bibnamefont {Modi}},\ }\href {https://doi.org/10.1103/PhysRevA.97.012127} {\bibfield  {journal} {\bibinfo  {journal} {Physical Review A}\ }\textbf {\bibinfo {volume} {97}},\ \bibinfo {pages} {012127} (\bibinfo {year} {2018}{\natexlab{a}})}\BibitemShut {NoStop}%
\bibitem [{\citenamefont {White}\ \emph {et~al.}(2022)\citenamefont {White}, \citenamefont {Pollock}, \citenamefont {Hollenberg}, \citenamefont {Modi},\ and\ \citenamefont {Hill}}]{White_PTT_22}%
  \BibitemOpen
  \bibfield  {author} {\bibinfo {author} {\bibfnamefont {G.}~\bibnamefont {White}}, \bibinfo {author} {\bibfnamefont {F.}~\bibnamefont {Pollock}}, \bibinfo {author} {\bibfnamefont {L.}~\bibnamefont {Hollenberg}}, \bibinfo {author} {\bibfnamefont {K.}~\bibnamefont {Modi}},\ and\ \bibinfo {author} {\bibfnamefont {C.}~\bibnamefont {Hill}},\ }\href {https://doi.org/10.1103/PRXQuantum.3.020344} {\bibfield  {journal} {\bibinfo  {journal} {PRX Quantum}\ }\textbf {\bibinfo {volume} {3}},\ \bibinfo {pages} {020344} (\bibinfo {year} {2022})}\BibitemShut {NoStop}%
\bibitem [{\citenamefont {Pollock}\ \emph {et~al.}(2018{\natexlab{b}})\citenamefont {Pollock}, \citenamefont {Rodr\'{\i}guez-Rosario}, \citenamefont {Frauenheim}, \citenamefont {Paternostro},\ and\ \citenamefont {Modi}}]{Pollock_PTMarkov}%
  \BibitemOpen
  \bibfield  {author} {\bibinfo {author} {\bibfnamefont {F.~A.}\ \bibnamefont {Pollock}}, \bibinfo {author} {\bibfnamefont {C.}~\bibnamefont {Rodr\'{\i}guez-Rosario}}, \bibinfo {author} {\bibfnamefont {T.}~\bibnamefont {Frauenheim}}, \bibinfo {author} {\bibfnamefont {M.}~\bibnamefont {Paternostro}},\ and\ \bibinfo {author} {\bibfnamefont {K.}~\bibnamefont {Modi}},\ }\href {https://doi.org/10.1103/PhysRevLett.120.040405} {\bibfield  {journal} {\bibinfo  {journal} {Phys. Rev. Lett.}\ }\textbf {\bibinfo {volume} {120}},\ \bibinfo {pages} {040405} (\bibinfo {year} {2018}{\natexlab{b}})}\BibitemShut {NoStop}%
\bibitem [{\citenamefont {Cygorek}\ and\ \citenamefont {Gauger}(2024)}]{cygorek2024understanding}%
  \BibitemOpen
  \bibfield  {author} {\bibinfo {author} {\bibfnamefont {M.}~\bibnamefont {Cygorek}}\ and\ \bibinfo {author} {\bibfnamefont {E.~M.}\ \bibnamefont {Gauger}},\ }\href@noop {} {\bibinfo {title} {Understanding and utilizing the inner bonds of process tensors}} (\bibinfo {year} {2024}),\ \Eprint {https://arxiv.org/abs/2404.01287} {arXiv:2404.01287 [quant-ph]} \BibitemShut {NoStop}%
\bibitem [{\citenamefont {Samach}\ \emph {et~al.}(2022)\citenamefont {Samach}, \citenamefont {Greene}, \citenamefont {Borregaard}, \citenamefont {Christandl}, \citenamefont {Barreto}, \citenamefont {Kim}, \citenamefont {McNally}, \citenamefont {Melville}, \citenamefont {Niedzielski}, \citenamefont {Sung}, \citenamefont {Rosenberg}, \citenamefont {Schwartz}, \citenamefont {Yoder}, \citenamefont {Orlando}, \citenamefont {Wang}, \citenamefont {Gustavsson}, \citenamefont {Kjaergaard},\ and\ \citenamefont {Oliver}}]{samach_lindblad_2022}%
  \BibitemOpen
  \bibfield  {author} {\bibinfo {author} {\bibfnamefont {G.~O.}\ \bibnamefont {Samach}}, \bibinfo {author} {\bibfnamefont {A.}~\bibnamefont {Greene}}, \bibinfo {author} {\bibfnamefont {J.}~\bibnamefont {Borregaard}}, \bibinfo {author} {\bibfnamefont {M.}~\bibnamefont {Christandl}}, \bibinfo {author} {\bibfnamefont {J.}~\bibnamefont {Barreto}}, \bibinfo {author} {\bibfnamefont {D.~K.}\ \bibnamefont {Kim}}, \bibinfo {author} {\bibfnamefont {C.~M.}\ \bibnamefont {McNally}}, \bibinfo {author} {\bibfnamefont {A.}~\bibnamefont {Melville}}, \bibinfo {author} {\bibfnamefont {B.~M.}\ \bibnamefont {Niedzielski}}, \bibinfo {author} {\bibfnamefont {Y.}~\bibnamefont {Sung}}, \bibinfo {author} {\bibfnamefont {D.}~\bibnamefont {Rosenberg}}, \bibinfo {author} {\bibfnamefont {M.~E.}\ \bibnamefont {Schwartz}}, \bibinfo {author} {\bibfnamefont {J.~L.}\ \bibnamefont {Yoder}}, \bibinfo {author} {\bibfnamefont {T.~P.}\ \bibnamefont {Orlando}}, \bibinfo {author} {\bibfnamefont {J.~I.-J.}\ \bibnamefont {Wang}}, \bibinfo {author}
  {\bibfnamefont {S.}~\bibnamefont {Gustavsson}}, \bibinfo {author} {\bibfnamefont {M.}~\bibnamefont {Kjaergaard}},\ and\ \bibinfo {author} {\bibfnamefont {W.~D.}\ \bibnamefont {Oliver}},\ }\href {https://doi.org/10.1103/PhysRevApplied.18.064056} {\bibfield  {journal} {\bibinfo  {journal} {Physical Review Applied}\ }\textbf {\bibinfo {volume} {18}},\ \bibinfo {pages} {064056} (\bibinfo {year} {2022})}\BibitemShut {NoStop}%
\bibitem [{\citenamefont {Koong}\ \emph {et~al.}(2022)\citenamefont {Koong}, \citenamefont {Cygorek}, \citenamefont {Scerri}, \citenamefont {Santana}, \citenamefont {{Suk In Park}}, \citenamefont {Song}, \citenamefont {Song}, \citenamefont {Gauger},\ and\ \citenamefont {Gerardot}}]{koong_coherence_2022}%
  \BibitemOpen
  \bibfield  {author} {\bibinfo {author} {\bibfnamefont {Z.~X.}\ \bibnamefont {Koong}}, \bibinfo {author} {\bibfnamefont {M.}~\bibnamefont {Cygorek}}, \bibinfo {author} {\bibfnamefont {D.}~\bibnamefont {Scerri}}, \bibinfo {author} {\bibfnamefont {T.~S.}\ \bibnamefont {Santana}}, \bibinfo {author} {\bibnamefont {{Suk In Park}}}, \bibinfo {author} {\bibfnamefont {J.~D.}\ \bibnamefont {Song}}, \bibinfo {author} {\bibfnamefont {J.~D.}\ \bibnamefont {Song}}, \bibinfo {author} {\bibfnamefont {E.~M.}\ \bibnamefont {Gauger}},\ and\ \bibinfo {author} {\bibfnamefont {B.~D.}\ \bibnamefont {Gerardot}},\ }\href {https://doi.org/10.1126/sciadv.abm8171} {\bibfield  {journal} {\bibinfo  {journal} {Science Advances}\ }\textbf {\bibinfo {volume} {8}},\ \bibinfo {pages} {eabm8171} (\bibinfo {year} {2022})}\BibitemShut {NoStop}%
\bibitem [{\citenamefont {Gorini}\ \emph {et~al.}(1976)\citenamefont {Gorini}, \citenamefont {Kossakowski}, \citenamefont {Kossakowski}, ,\ and\ \citenamefont {Sudarshan}}]{vittorio_gorini_completely_1976}%
  \BibitemOpen
  \bibfield  {author} {\bibinfo {author} {\bibfnamefont {V.}~\bibnamefont {Gorini}}, \bibinfo {author} {\bibfnamefont {A.}~\bibnamefont {Kossakowski}}, \bibinfo {author} {\bibfnamefont {A.}~\bibnamefont {Kossakowski}}, ,\ and\ \bibinfo {author} {\bibfnamefont {E.~C.~G.}\ \bibnamefont {Sudarshan}},\ }\href {https://doi.org/10.1063/1.522979} {\bibfield  {journal} {\bibinfo  {journal} {Journal of Mathematical Physics}\ }\textbf {\bibinfo {volume} {17}},\ \bibinfo {pages} {821} (\bibinfo {year} {1976})}\BibitemShut {NoStop}%
\bibitem [{\citenamefont {Lindblad}(1976)}]{lindblad_generators_1976}%
  \BibitemOpen
  \bibfield  {author} {\bibinfo {author} {\bibfnamefont {G.}~\bibnamefont {Lindblad}},\ }\href {https://doi.org/10.1007/BF01608499} {\bibfield  {journal} {\bibinfo  {journal} {Communications in Mathematical Physics}\ }\textbf {\bibinfo {volume} {48}},\ \bibinfo {pages} {119} (\bibinfo {year} {1976})}\BibitemShut {NoStop}%
\bibitem [{\citenamefont {Gammelmark}\ and\ \citenamefont {Mølmer}(2013)}]{gammelmark_bayesian_2013}%
  \BibitemOpen
  \bibfield  {author} {\bibinfo {author} {\bibfnamefont {S.}~\bibnamefont {Gammelmark}}\ and\ \bibinfo {author} {\bibfnamefont {K.}~\bibnamefont {Mølmer}},\ }\href {https://doi.org/10.1103/PhysRevA.87.032115} {\bibfield  {journal} {\bibinfo  {journal} {Physical Review A}\ }\textbf {\bibinfo {volume} {87}},\ \bibinfo {pages} {032115} (\bibinfo {year} {2013})}\BibitemShut {NoStop}%
\bibitem [{\citenamefont {Assmann}\ \emph {et~al.}(2010)\citenamefont {Assmann}, \citenamefont {Veit}, , \citenamefont {Tempel}, \citenamefont {Berstermann}, \citenamefont {Stolz}, \citenamefont {van~der Poel}, \citenamefont {Hvam},\ and\ \citenamefont {Bayer}}]{marc_asmann_measuring_2010}%
  \BibitemOpen
  \bibfield  {author} {\bibinfo {author} {\bibfnamefont {M.}~\bibnamefont {Assmann}}, \bibinfo {author} {\bibfnamefont {F.}~\bibnamefont {Veit}}, , \bibinfo {author} {\bibfnamefont {J.-S.}\ \bibnamefont {Tempel}}, \bibinfo {author} {\bibfnamefont {T.}~\bibnamefont {Berstermann}}, \bibinfo {author} {\bibfnamefont {H.}~\bibnamefont {Stolz}}, \bibinfo {author} {\bibfnamefont {M.}~\bibnamefont {van~der Poel}}, \bibinfo {author} {\bibfnamefont {J.~M.}\ \bibnamefont {Hvam}},\ and\ \bibinfo {author} {\bibfnamefont {M.}~\bibnamefont {Bayer}},\ }\href {https://doi.org/10.1364\oe.18.020229} {\bibfield  {journal} {\bibinfo  {journal} {Optics Express}\ }\textbf {\bibinfo {volume} {18}},\ \bibinfo {pages} {20229} (\bibinfo {year} {2010})}\BibitemShut {NoStop}%
\bibitem [{\citenamefont {Kim}\ \emph {et~al.}(2018)\citenamefont {Kim}, \citenamefont {Aghaeimeibodi}, \citenamefont {Richardson}, \citenamefont {Leavitt},\ and\ \citenamefont {Waks}}]{kim_super-radiant_2018}%
  \BibitemOpen
  \bibfield  {author} {\bibinfo {author} {\bibfnamefont {J.-H.}\ \bibnamefont {Kim}}, \bibinfo {author} {\bibfnamefont {S.}~\bibnamefont {Aghaeimeibodi}}, \bibinfo {author} {\bibfnamefont {C.~J.~K.}\ \bibnamefont {Richardson}}, \bibinfo {author} {\bibfnamefont {R.~P.}\ \bibnamefont {Leavitt}},\ and\ \bibinfo {author} {\bibfnamefont {E.}~\bibnamefont {Waks}},\ }\href {https://doi.org/10.1021/acs.nanolett.8b01133} {\bibfield  {journal} {\bibinfo  {journal} {Nano Letters}\ }\textbf {\bibinfo {volume} {18}},\ \bibinfo {pages} {4734} (\bibinfo {year} {2018})}\BibitemShut {NoStop}%
\bibitem [{\citenamefont {{D. Levonian}}\ \emph {et~al.}(2021)\citenamefont {{D. Levonian}}, \citenamefont {{R. Riedinger}}, \citenamefont {{B. Machielse}}, \citenamefont {{E. Knall}}, \citenamefont {{M. Bhaskar}}, \citenamefont {{C. Knaut}}, \citenamefont {{R. Bekenstein}}, \citenamefont {{Hongkun Park}}, \citenamefont {{M. Lončar}},\ and\ \citenamefont {{M. Lukin}}}]{d_levonian_optical_2021}%
  \BibitemOpen
  \bibfield  {author} {\bibinfo {author} {\bibnamefont {{D. Levonian}}}, \bibinfo {author} {\bibnamefont {{R. Riedinger}}}, \bibinfo {author} {\bibnamefont {{B. Machielse}}}, \bibinfo {author} {\bibnamefont {{E. Knall}}}, \bibinfo {author} {\bibnamefont {{M. Bhaskar}}}, \bibinfo {author} {\bibnamefont {{C. Knaut}}}, \bibinfo {author} {\bibnamefont {{R. Bekenstein}}}, \bibinfo {author} {\bibnamefont {{Hongkun Park}}}, \bibinfo {author} {\bibnamefont {{M. Lončar}}},\ and\ \bibinfo {author} {\bibnamefont {{M. Lukin}}},\ }\bibfield  {journal} {\bibinfo  {journal} {Physical Review Letters}\ }\href {https://doi.org/10.1103/physrevlett.128.213602} {10.1103/physrevlett.128.213602} (\bibinfo {year} {2021})\BibitemShut {NoStop}%
\bibitem [{\citenamefont {{Rebecca E. K. Fishman}}\ \emph {et~al.}(2023)\citenamefont {{Rebecca E. K. Fishman}}, \citenamefont {{Raj N. Patel}}, \citenamefont {{David A. Hopper}}, \citenamefont {{Tzu-Yung Huang}},\ and\ \citenamefont {{Lee C. Bassett}}}]{rebecca_e_k_fishman_photon-emission-correlation_2023}%
  \BibitemOpen
  \bibfield  {author} {\bibinfo {author} {\bibnamefont {{Rebecca E. K. Fishman}}}, \bibinfo {author} {\bibnamefont {{Raj N. Patel}}}, \bibinfo {author} {\bibnamefont {{David A. Hopper}}}, \bibinfo {author} {\bibnamefont {{Tzu-Yung Huang}}},\ and\ \bibinfo {author} {\bibnamefont {{Lee C. Bassett}}},\ }\bibfield  {journal} {\bibinfo  {journal} {PRX Quantum}\ }\textbf {\bibinfo {volume} {4}},\ \href {https://doi.org/10.1103/prxquantum.4.010202} {10.1103/prxquantum.4.010202} (\bibinfo {year} {2023})\BibitemShut {NoStop}%
\bibitem [{\citenamefont {Cygorek}\ \emph {et~al.}(2023)\citenamefont {Cygorek}, \citenamefont {Scerri}, \citenamefont {Santana}, \citenamefont {Koong}, \citenamefont {Gerardot},\ and\ \citenamefont {Gauger}}]{cygorek_signatures_2023}%
  \BibitemOpen
  \bibfield  {author} {\bibinfo {author} {\bibfnamefont {M.}~\bibnamefont {Cygorek}}, \bibinfo {author} {\bibfnamefont {E.~D.}\ \bibnamefont {Scerri}}, \bibinfo {author} {\bibfnamefont {T.~S.}\ \bibnamefont {Santana}}, \bibinfo {author} {\bibfnamefont {Z.~X.}\ \bibnamefont {Koong}}, \bibinfo {author} {\bibfnamefont {B.~D.}\ \bibnamefont {Gerardot}},\ and\ \bibinfo {author} {\bibfnamefont {E.~M.}\ \bibnamefont {Gauger}},\ }\href {https://doi.org/10.1103/PhysRevA.107.023718} {\bibfield  {journal} {\bibinfo  {journal} {Physical Review A}\ }\textbf {\bibinfo {volume} {107}},\ \bibinfo {pages} {023718} (\bibinfo {year} {2023})}\BibitemShut {NoStop}%
\bibitem [{\citenamefont {{Daniil M. Lukin}}\ \emph {et~al.}(2023)\citenamefont {{Daniil M. Lukin}}, \citenamefont {{Melissa A. Guidry}}, \citenamefont {{Joshua Yang}}, \citenamefont {{Misagh Ghezellou}}, \citenamefont {{Sattwik Deb Mishra}}, \citenamefont {{Hiroshi Abe}}, \citenamefont {{Takeshi Ohshima}}, \citenamefont {{Jawad Ul-Hassan}},\ and\ \citenamefont {{Jelena Vučković}}}]{daniil_m_lukin_two-emitter_2023}%
  \BibitemOpen
  \bibfield  {author} {\bibinfo {author} {\bibnamefont {{Daniil M. Lukin}}}, \bibinfo {author} {\bibnamefont {{Melissa A. Guidry}}}, \bibinfo {author} {\bibnamefont {{Joshua Yang}}}, \bibinfo {author} {\bibnamefont {{Misagh Ghezellou}}}, \bibinfo {author} {\bibnamefont {{Sattwik Deb Mishra}}}, \bibinfo {author} {\bibnamefont {{Hiroshi Abe}}}, \bibinfo {author} {\bibnamefont {{Takeshi Ohshima}}}, \bibinfo {author} {\bibnamefont {{Jawad Ul-Hassan}}},\ and\ \bibinfo {author} {\bibnamefont {{Jelena Vučković}}},\ }\bibfield  {journal} {\bibinfo  {journal} {Physical Review X}\ }\textbf {\bibinfo {volume} {13}},\ \href {https://doi.org/10.1103/physrevx.13.011005} {10.1103/physrevx.13.011005} (\bibinfo {year} {2023})\BibitemShut {NoStop}%
\bibitem [{\citenamefont {Tiranov}\ \emph {et~al.}(2023)\citenamefont {Tiranov}, \citenamefont {Angelopoulou}, \citenamefont {Van~Diepen}, \citenamefont {Schrinski}, \citenamefont {Sandberg}, \citenamefont {Wang}, \citenamefont {Midolo}, \citenamefont {Scholz}, \citenamefont {Wieck}, \citenamefont {Ludwig}, \citenamefont {Sørensen},\ and\ \citenamefont {Lodahl}}]{tiranov_collective_2023}%
  \BibitemOpen
  \bibfield  {author} {\bibinfo {author} {\bibfnamefont {A.}~\bibnamefont {Tiranov}}, \bibinfo {author} {\bibfnamefont {V.}~\bibnamefont {Angelopoulou}}, \bibinfo {author} {\bibfnamefont {C.~J.}\ \bibnamefont {Van~Diepen}}, \bibinfo {author} {\bibfnamefont {B.}~\bibnamefont {Schrinski}}, \bibinfo {author} {\bibfnamefont {O.~A.~D.}\ \bibnamefont {Sandberg}}, \bibinfo {author} {\bibfnamefont {Y.}~\bibnamefont {Wang}}, \bibinfo {author} {\bibfnamefont {L.}~\bibnamefont {Midolo}}, \bibinfo {author} {\bibfnamefont {S.}~\bibnamefont {Scholz}}, \bibinfo {author} {\bibfnamefont {A.~D.}\ \bibnamefont {Wieck}}, \bibinfo {author} {\bibfnamefont {A.}~\bibnamefont {Ludwig}}, \bibinfo {author} {\bibfnamefont {A.~S.}\ \bibnamefont {Sørensen}},\ and\ \bibinfo {author} {\bibfnamefont {P.}~\bibnamefont {Lodahl}},\ }\href {https://doi.org/10.1126/science.ade9324} {\bibfield  {journal} {\bibinfo  {journal} {Science}\ }\textbf {\bibinfo {volume} {379}},\ \bibinfo {pages} {389} (\bibinfo {year} {2023})}\BibitemShut {NoStop}%
\bibitem [{\citenamefont {Banchi}\ \emph {et~al.}(2018)\citenamefont {Banchi}, \citenamefont {Grant}, \citenamefont {Rocchetto},\ and\ \citenamefont {Severini}}]{banchi_modelling_2018}%
  \BibitemOpen
  \bibfield  {author} {\bibinfo {author} {\bibfnamefont {L.}~\bibnamefont {Banchi}}, \bibinfo {author} {\bibfnamefont {E.}~\bibnamefont {Grant}}, \bibinfo {author} {\bibfnamefont {A.}~\bibnamefont {Rocchetto}},\ and\ \bibinfo {author} {\bibfnamefont {S.}~\bibnamefont {Severini}},\ }\href {https://doi.org/10.1088/1367-2630/aaf749} {\bibfield  {journal} {\bibinfo  {journal} {New Journal of Physics}\ }\textbf {\bibinfo {volume} {20}},\ \bibinfo {pages} {123030} (\bibinfo {year} {2018})}\BibitemShut {NoStop}%
\bibitem [{\citenamefont {Hartmann}\ and\ \citenamefont {Carleo}(2019)}]{hartmann_neural-network_2019}%
  \BibitemOpen
  \bibfield  {author} {\bibinfo {author} {\bibfnamefont {M.~J.}\ \bibnamefont {Hartmann}}\ and\ \bibinfo {author} {\bibfnamefont {G.}~\bibnamefont {Carleo}},\ }\href {https://doi.org/10.1103/PhysRevLett.122.250502} {\bibfield  {journal} {\bibinfo  {journal} {Physical Review Letters}\ }\textbf {\bibinfo {volume} {122}},\ \bibinfo {pages} {250502} (\bibinfo {year} {2019})}\BibitemShut {NoStop}%
\bibitem [{\citenamefont {Luchnikov}\ \emph {et~al.}(2020)\citenamefont {Luchnikov}, \citenamefont {Vintskevich}, \citenamefont {Grigoriev},\ and\ \citenamefont {Filippov}}]{luchnikov_machine_2020}%
  \BibitemOpen
  \bibfield  {author} {\bibinfo {author} {\bibfnamefont {I.}~\bibnamefont {Luchnikov}}, \bibinfo {author} {\bibfnamefont {S.}~\bibnamefont {Vintskevich}}, \bibinfo {author} {\bibfnamefont {D.}~\bibnamefont {Grigoriev}},\ and\ \bibinfo {author} {\bibfnamefont {S.}~\bibnamefont {Filippov}},\ }\href {https://doi.org/10.1103/PhysRevLett.124.140502} {\bibfield  {journal} {\bibinfo  {journal} {Physical Review Letters}\ }\textbf {\bibinfo {volume} {124}},\ \bibinfo {pages} {140502} (\bibinfo {year} {2020})}\BibitemShut {NoStop}%
\bibitem [{\citenamefont {Mazza}\ \emph {et~al.}(2021)\citenamefont {Mazza}, \citenamefont {Zietlow}, \citenamefont {Carollo}, \citenamefont {Andergassen}, \citenamefont {Martius},\ and\ \citenamefont {Lesanovsky}}]{mazza_machine_2021}%
  \BibitemOpen
  \bibfield  {author} {\bibinfo {author} {\bibfnamefont {P.~P.}\ \bibnamefont {Mazza}}, \bibinfo {author} {\bibfnamefont {D.}~\bibnamefont {Zietlow}}, \bibinfo {author} {\bibfnamefont {F.}~\bibnamefont {Carollo}}, \bibinfo {author} {\bibfnamefont {S.}~\bibnamefont {Andergassen}}, \bibinfo {author} {\bibfnamefont {G.}~\bibnamefont {Martius}},\ and\ \bibinfo {author} {\bibfnamefont {I.}~\bibnamefont {Lesanovsky}},\ }\href {http://arxiv.org/abs/2101.08591} {\bibfield  {journal} {\bibinfo  {journal} {arXiv:2101.08591 [cond-mat, physics:quant-ph]}\ } (\bibinfo {year} {2021})}\BibitemShut {NoStop}%
\bibitem [{\citenamefont {Krastanov}\ \emph {et~al.}(2020)\citenamefont {Krastanov}, \citenamefont {Head-Marsden}, \citenamefont {Zhou}, \citenamefont {Flammia}, \citenamefont {Jiang},\ and\ \citenamefont {Narang}}]{krastanov_unboxing_2020}%
  \BibitemOpen
  \bibfield  {author} {\bibinfo {author} {\bibfnamefont {S.}~\bibnamefont {Krastanov}}, \bibinfo {author} {\bibfnamefont {K.}~\bibnamefont {Head-Marsden}}, \bibinfo {author} {\bibfnamefont {S.}~\bibnamefont {Zhou}}, \bibinfo {author} {\bibfnamefont {S.~T.}\ \bibnamefont {Flammia}}, \bibinfo {author} {\bibfnamefont {L.}~\bibnamefont {Jiang}},\ and\ \bibinfo {author} {\bibfnamefont {P.}~\bibnamefont {Narang}},\ }\href {http://arxiv.org/abs/2009.03902} {\bibfield  {journal} {\bibinfo  {journal} {arXiv:2009.03902 [quant-ph]}\ } (\bibinfo {year} {2020})}\BibitemShut {NoStop}%
\bibitem [{\citenamefont {Carnazza}\ \emph {et~al.}(2022)\citenamefont {Carnazza}, \citenamefont {Carollo}, \citenamefont {Zietlow}, \citenamefont {Andergassen}, \citenamefont {Martius},\ and\ \citenamefont {Lesanovsky}}]{carnazza_inferring_2022}%
  \BibitemOpen
  \bibfield  {author} {\bibinfo {author} {\bibfnamefont {F.}~\bibnamefont {Carnazza}}, \bibinfo {author} {\bibfnamefont {F.}~\bibnamefont {Carollo}}, \bibinfo {author} {\bibfnamefont {D.}~\bibnamefont {Zietlow}}, \bibinfo {author} {\bibfnamefont {S.}~\bibnamefont {Andergassen}}, \bibinfo {author} {\bibfnamefont {G.}~\bibnamefont {Martius}},\ and\ \bibinfo {author} {\bibfnamefont {I.}~\bibnamefont {Lesanovsky}},\ }\href {https://doi.org/10.1088/1367-2630/ac7df6} {\bibfield  {journal} {\bibinfo  {journal} {New Journal of Physics}\ }\textbf {\bibinfo {volume} {24}},\ \bibinfo {pages} {073033} (\bibinfo {year} {2022})}\BibitemShut {NoStop}%
\bibitem [{\citenamefont {Cemin}\ \emph {et~al.}(2024)\citenamefont {Cemin}, \citenamefont {Carnazza}, \citenamefont {Andergassen}, \citenamefont {Martius}, \citenamefont {Carollo},\ and\ \citenamefont {Lesanovsky}}]{cemin_inferring_2024}%
  \BibitemOpen
  \bibfield  {author} {\bibinfo {author} {\bibfnamefont {G.}~\bibnamefont {Cemin}}, \bibinfo {author} {\bibfnamefont {F.}~\bibnamefont {Carnazza}}, \bibinfo {author} {\bibfnamefont {S.}~\bibnamefont {Andergassen}}, \bibinfo {author} {\bibfnamefont {G.}~\bibnamefont {Martius}}, \bibinfo {author} {\bibfnamefont {F.}~\bibnamefont {Carollo}},\ and\ \bibinfo {author} {\bibfnamefont {I.}~\bibnamefont {Lesanovsky}},\ }\href {https://doi.org/10.1103/PhysRevApplied.21.L041001} {\bibfield  {journal} {\bibinfo  {journal} {Physical Review Applied}\ }\textbf {\bibinfo {volume} {21}},\ \bibinfo {pages} {L041001} (\bibinfo {year} {2024})}\BibitemShut {NoStop}%
\bibitem [{\citenamefont {Cranmer}\ \emph {et~al.}(2020)\citenamefont {Cranmer}, \citenamefont {Sanchez-Gonzalez}, , \citenamefont {Battaglia}, \citenamefont {{Rui Xu}}, \citenamefont {Xu}, \citenamefont {Cranmer}, \citenamefont {Spergel},\ and\ \citenamefont {Ho}}]{miles_cranmer_discovering_2020}%
  \BibitemOpen
  \bibfield  {author} {\bibinfo {author} {\bibfnamefont {M.}~\bibnamefont {Cranmer}}, \bibinfo {author} {\bibfnamefont {A.}~\bibnamefont {Sanchez-Gonzalez}}, , \bibinfo {author} {\bibfnamefont {P.~W.}\ \bibnamefont {Battaglia}}, \bibinfo {author} {\bibnamefont {{Rui Xu}}}, \bibinfo {author} {\bibfnamefont {R.}~\bibnamefont {Xu}}, \bibinfo {author} {\bibfnamefont {K.}~\bibnamefont {Cranmer}}, \bibinfo {author} {\bibfnamefont {D.~N.}\ \bibnamefont {Spergel}},\ and\ \bibinfo {author} {\bibfnamefont {S.}~\bibnamefont {Ho}},\ }\href@noop {} {\bibfield  {journal} {\bibinfo  {journal} {Neural Information Processing Systems}\ } (\bibinfo {year} {2020})}\BibitemShut {NoStop}%
\bibitem [{\citenamefont {Nautrup}\ \emph {et~al.}(2022)\citenamefont {Nautrup}, \citenamefont {Metger}, \citenamefont {Iten}, \citenamefont {Jerbi}, \citenamefont {Trenkwalder}, \citenamefont {Wilming}, \citenamefont {Briegel},\ and\ \citenamefont {Renner}}]{nautrup_operationally_2022}%
  \BibitemOpen
  \bibfield  {author} {\bibinfo {author} {\bibfnamefont {H.~P.}\ \bibnamefont {Nautrup}}, \bibinfo {author} {\bibfnamefont {T.}~\bibnamefont {Metger}}, \bibinfo {author} {\bibfnamefont {R.}~\bibnamefont {Iten}}, \bibinfo {author} {\bibfnamefont {S.}~\bibnamefont {Jerbi}}, \bibinfo {author} {\bibfnamefont {L.~M.}\ \bibnamefont {Trenkwalder}}, \bibinfo {author} {\bibfnamefont {H.}~\bibnamefont {Wilming}}, \bibinfo {author} {\bibfnamefont {H.~J.}\ \bibnamefont {Briegel}},\ and\ \bibinfo {author} {\bibfnamefont {R.}~\bibnamefont {Renner}},\ }\href {https://doi.org/10.1088/2632-2153/ac9ae8} {\bibfield  {journal} {\bibinfo  {journal} {Machine Learning: Science and Technology}\ }\textbf {\bibinfo {volume} {3}},\ \bibinfo {pages} {045025} (\bibinfo {year} {2022})}\BibitemShut {NoStop}%
\bibitem [{\citenamefont {Craig}\ \emph {et~al.}(2024{\natexlab{b}})\citenamefont {Craig}, \citenamefont {Ares},\ and\ \citenamefont {Gauger}}]{Craig2024}%
  \BibitemOpen
  \bibfield  {author} {\bibinfo {author} {\bibfnamefont {D.~L.}\ \bibnamefont {Craig}}, \bibinfo {author} {\bibfnamefont {N.}~\bibnamefont {Ares}},\ and\ \bibinfo {author} {\bibfnamefont {E.~M.}\ \bibnamefont {Gauger}},\ }\href {https://arxiv.org/abs/2403.04678} {\bibfield  {journal} {\bibinfo  {journal} {Physical Review Research}\ } (\bibinfo {year} {2024}{\natexlab{b}})}\BibitemShut {NoStop}%
\bibitem [{\citenamefont {Wiercinski}\ \emph {et~al.}(2023)\citenamefont {Wiercinski}, \citenamefont {Gauger},\ and\ \citenamefont {Cygorek}}]{wiercinski_phonon_2023}%
  \BibitemOpen
  \bibfield  {author} {\bibinfo {author} {\bibfnamefont {J.}~\bibnamefont {Wiercinski}}, \bibinfo {author} {\bibfnamefont {E.~M.}\ \bibnamefont {Gauger}},\ and\ \bibinfo {author} {\bibfnamefont {M.}~\bibnamefont {Cygorek}},\ }\href {https://doi.org/10.1103/PhysRevResearch.5.013176} {\bibfield  {journal} {\bibinfo  {journal} {Physical Review Research}\ }\textbf {\bibinfo {volume} {5}},\ \bibinfo {pages} {013176} (\bibinfo {year} {2023})}\BibitemShut {NoStop}%
\bibitem [{\citenamefont {Auffèves}\ \emph {et~al.}(2011)\citenamefont {Auffèves}, \citenamefont {Gerace}, \citenamefont {Portolan}, \citenamefont {Drezet},\ and\ \citenamefont {Santos}}]{auffeves_few_2011}%
  \BibitemOpen
  \bibfield  {author} {\bibinfo {author} {\bibfnamefont {A.}~\bibnamefont {Auffèves}}, \bibinfo {author} {\bibfnamefont {D.}~\bibnamefont {Gerace}}, \bibinfo {author} {\bibfnamefont {S.}~\bibnamefont {Portolan}}, \bibinfo {author} {\bibfnamefont {A.}~\bibnamefont {Drezet}},\ and\ \bibinfo {author} {\bibfnamefont {M.~F.}\ \bibnamefont {Santos}},\ }\href {https://doi.org/10.1088/1367-2630/13/9/093020} {\bibfield  {journal} {\bibinfo  {journal} {New Journal of Physics}\ }\textbf {\bibinfo {volume} {13}},\ \bibinfo {pages} {093020} (\bibinfo {year} {2011})}\BibitemShut {NoStop}%
\bibitem [{\citenamefont {Green}\ and\ \citenamefont {Green}(1995)}]{peter_j_green_reversible_1995}%
  \BibitemOpen
  \bibfield  {author} {\bibinfo {author} {\bibfnamefont {P.~J.}\ \bibnamefont {Green}}\ and\ \bibinfo {author} {\bibfnamefont {P.~H.}\ \bibnamefont {Green}},\ }\href {https://doi.org/10.1093/biomet/82.4.711} {\bibfield  {journal} {\bibinfo  {journal} {Biometrika}\ }\textbf {\bibinfo {volume} {82}},\ \bibinfo {pages} {711} (\bibinfo {year} {1995})}\BibitemShut {NoStop}%
\bibitem [{Note1()}]{Note1}%
  \BibitemOpen
  \bibinfo {note} {The identity is possible for a single 2LS only, when allowing operator complexity up to a fixed $\protect \mathcal {C}$ (in this work $\protect \mathcal {C}=2$). Its appearance increases confidence in $\protect \mathcal {M}_{1A}$ being the top model while also demonstrating our approach does not penalize additional processes to the extent that might suppress their inclusion if relevant.}\BibitemShut {Stop}%
\bibitem [{Note2()}]{Note2}%
  \BibitemOpen
  \bibinfo {note} {To be explicit, we have used the short notation $(\sigma _1^- + \sigma _2^-)/\protect \sqrt {2} \equiv (\sigma _1^- \otimes \protect \mathcal {I}_2 + \protect \mathcal {I}_1 \otimes \sigma _2^-)/\protect \sqrt {2}$.}\BibitemShut {Stop}%
\bibitem [{Note3()}]{Note3}%
  \BibitemOpen
  \bibinfo {note} {It might be interesting to explore dynamic or adaptive relative weighting of different types of input data as a means to enable both broad exploration and then optimisation in separate stages}\BibitemShut {NoStop}%
\bibitem [{Note4()}]{Note4}%
  \BibitemOpen
  \bibinfo {note} {Note that our choice of penalising a large number of terms is in spirit related to the idea of incentivising a low rank for the Kossakowski matrix, however, our choice of operator basis means there is no direct correspondence between that rank and the number of terms in our model.}\BibitemShut {Stop}%
\bibitem [{\citenamefont {Rosenberg}\ \emph {et~al.}(2025)\citenamefont {Rosenberg}, \citenamefont {Kuffer}, \citenamefont {Zohar}, \citenamefont {Stöhr}, \citenamefont {Denisenko}, \citenamefont {Zwick}, \citenamefont {Álvarez},\ and\ \citenamefont {Finkler}}]{rosenberg_witnessing_2025}%
  \BibitemOpen
  \bibfield  {author} {\bibinfo {author} {\bibfnamefont {J.~W.}\ \bibnamefont {Rosenberg}}, \bibinfo {author} {\bibfnamefont {M.}~\bibnamefont {Kuffer}}, \bibinfo {author} {\bibfnamefont {I.}~\bibnamefont {Zohar}}, \bibinfo {author} {\bibfnamefont {R.}~\bibnamefont {Stöhr}}, \bibinfo {author} {\bibfnamefont {A.}~\bibnamefont {Denisenko}}, \bibinfo {author} {\bibfnamefont {A.}~\bibnamefont {Zwick}}, \bibinfo {author} {\bibfnamefont {G.~A.}\ \bibnamefont {Álvarez}},\ and\ \bibinfo {author} {\bibfnamefont {A.}~\bibnamefont {Finkler}},\ }\href {https://doi.org/10.48550/arXiv.2501.05814} {\bibinfo {title} {Witnessing non-stationary and non-{Markovian} environments with a quantum sensor}} (\bibinfo {year} {2025}),\ \bibinfo {note} {arXiv:2501.05814}\BibitemShut {NoStop}%
\bibitem [{\citenamefont {Bonato}\ \emph {et~al.}(2016)\citenamefont {Bonato}, \citenamefont {Blok}, \citenamefont {Dinani}, \citenamefont {Berry}, \citenamefont {Markham}, \citenamefont {Twitchen},\ and\ \citenamefont {Hanson}}]{bonato_optimized_2016}%
  \BibitemOpen
  \bibfield  {author} {\bibinfo {author} {\bibfnamefont {C.}~\bibnamefont {Bonato}}, \bibinfo {author} {\bibfnamefont {M.~S.}\ \bibnamefont {Blok}}, \bibinfo {author} {\bibfnamefont {H.~T.}\ \bibnamefont {Dinani}}, \bibinfo {author} {\bibfnamefont {D.~W.}\ \bibnamefont {Berry}}, \bibinfo {author} {\bibfnamefont {M.~L.}\ \bibnamefont {Markham}}, \bibinfo {author} {\bibfnamefont {D.~J.}\ \bibnamefont {Twitchen}},\ and\ \bibinfo {author} {\bibfnamefont {R.}~\bibnamefont {Hanson}},\ }\href {https://doi.org/10.1038/nnano.2015.261} {\bibfield  {journal} {\bibinfo  {journal} {Nature Nanotechnology}\ }\textbf {\bibinfo {volume} {11}},\ \bibinfo {pages} {247} (\bibinfo {year} {2016})}\BibitemShut {NoStop}%
\bibitem [{\citenamefont {Arshad}\ \emph {et~al.}(2024)\citenamefont {Arshad}, \citenamefont {Bekker}, \citenamefont {Haylock}, \citenamefont {Skrzypczak}, \citenamefont {White}, \citenamefont {Griffiths}, \citenamefont {Gore}, \citenamefont {Morley}, \citenamefont {Salter}, \citenamefont {Smith}, \citenamefont {Zohar}, \citenamefont {Finkler}, \citenamefont {Altmann}, \citenamefont {Gauger},\ and\ \citenamefont {Bonato}}]{arshad_real-time_2024}%
  \BibitemOpen
  \bibfield  {author} {\bibinfo {author} {\bibfnamefont {M.~J.}\ \bibnamefont {Arshad}}, \bibinfo {author} {\bibfnamefont {C.}~\bibnamefont {Bekker}}, \bibinfo {author} {\bibfnamefont {B.}~\bibnamefont {Haylock}}, \bibinfo {author} {\bibfnamefont {K.}~\bibnamefont {Skrzypczak}}, \bibinfo {author} {\bibfnamefont {D.}~\bibnamefont {White}}, \bibinfo {author} {\bibfnamefont {B.}~\bibnamefont {Griffiths}}, \bibinfo {author} {\bibfnamefont {J.}~\bibnamefont {Gore}}, \bibinfo {author} {\bibfnamefont {G.~W.}\ \bibnamefont {Morley}}, \bibinfo {author} {\bibfnamefont {P.}~\bibnamefont {Salter}}, \bibinfo {author} {\bibfnamefont {J.}~\bibnamefont {Smith}}, \bibinfo {author} {\bibfnamefont {I.}~\bibnamefont {Zohar}}, \bibinfo {author} {\bibfnamefont {A.}~\bibnamefont {Finkler}}, \bibinfo {author} {\bibfnamefont {Y.}~\bibnamefont {Altmann}}, \bibinfo {author} {\bibfnamefont {E.~M.}\ \bibnamefont {Gauger}},\ and\ \bibinfo {author} {\bibfnamefont {C.}~\bibnamefont {Bonato}},\ }\href
  {https://doi.org/10.1103/PhysRevApplied.21.024026} {\bibfield  {journal} {\bibinfo  {journal} {Physical Review Applied}\ }\textbf {\bibinfo {volume} {21}},\ \bibinfo {pages} {024026} (\bibinfo {year} {2024})}\BibitemShut {NoStop}%
\bibitem [{\citenamefont {Berritta}\ \emph {et~al.}(2024)\citenamefont {Berritta}, \citenamefont {Krzywda}, \citenamefont {Benestad}, \citenamefont {van~der Heijden}, \citenamefont {Fedele}, \citenamefont {Fallahi}, \citenamefont {Gardner}, \citenamefont {Manfra}, \citenamefont {van Nieuwenburg}, \citenamefont {Danon}, \citenamefont {Chatterjee},\ and\ \citenamefont {Kuemmeth}}]{berritta_physics-informed_2024}%
  \BibitemOpen
  \bibfield  {author} {\bibinfo {author} {\bibfnamefont {F.}~\bibnamefont {Berritta}}, \bibinfo {author} {\bibfnamefont {J.~A.}\ \bibnamefont {Krzywda}}, \bibinfo {author} {\bibfnamefont {J.}~\bibnamefont {Benestad}}, \bibinfo {author} {\bibfnamefont {J.}~\bibnamefont {van~der Heijden}}, \bibinfo {author} {\bibfnamefont {F.}~\bibnamefont {Fedele}}, \bibinfo {author} {\bibfnamefont {S.}~\bibnamefont {Fallahi}}, \bibinfo {author} {\bibfnamefont {G.~C.}\ \bibnamefont {Gardner}}, \bibinfo {author} {\bibfnamefont {M.~J.}\ \bibnamefont {Manfra}}, \bibinfo {author} {\bibfnamefont {E.}~\bibnamefont {van Nieuwenburg}}, \bibinfo {author} {\bibfnamefont {J.}~\bibnamefont {Danon}}, \bibinfo {author} {\bibfnamefont {A.}~\bibnamefont {Chatterjee}},\ and\ \bibinfo {author} {\bibfnamefont {F.}~\bibnamefont {Kuemmeth}},\ }\href {https://doi.org/10.1103/PhysRevApplied.22.014033} {\bibfield  {journal} {\bibinfo  {journal} {Physical Review Applied}\ }\textbf {\bibinfo {volume} {22}},\ \bibinfo {pages} {014033} (\bibinfo {year}
  {2024})}\BibitemShut {NoStop}%
\bibitem [{\citenamefont {Belliardo}\ \emph {et~al.}(0)\citenamefont {Belliardo}, \citenamefont {Zoratti},\ and\ \citenamefont {Giovannetti}}]{belliardo_application_2024}%
  \BibitemOpen
  \bibfield  {author} {\bibinfo {author} {\bibfnamefont {F.}~\bibnamefont {Belliardo}}, \bibinfo {author} {\bibfnamefont {F.}~\bibnamefont {Zoratti}},\ and\ \bibinfo {author} {\bibfnamefont {V.}~\bibnamefont {Giovannetti}},\ }\href {https://doi.org/10.1142/S0219749924500023} {\bibfield  {journal} {\bibinfo  {journal} {International Journal of Quantum Information}\ }\textbf {\bibinfo {volume} {0}},\ \bibinfo {pages} {2450002} (\bibinfo {year} {0})},\ \Eprint {https://arxiv.org/abs/https://doi.org/10.1142/S0219749924500023} {https://doi.org/10.1142/S0219749924500023} \BibitemShut {NoStop}%
\bibitem [{\citenamefont {Sarra}\ and\ \citenamefont {Marquardt}(2023)}]{sarra_deep_2023}%
  \BibitemOpen
  \bibfield  {author} {\bibinfo {author} {\bibfnamefont {L.}~\bibnamefont {Sarra}}\ and\ \bibinfo {author} {\bibfnamefont {F.}~\bibnamefont {Marquardt}},\ }\href {https://doi.org/10.1088/2632-2153/ad020d} {\bibfield  {journal} {\bibinfo  {journal} {Machine Learning: Science and Technology}\ }\textbf {\bibinfo {volume} {4}},\ \bibinfo {pages} {045022} (\bibinfo {year} {2023})}\BibitemShut {NoStop}%
\bibitem [{\citenamefont {Schlimgen}\ \emph {et~al.}(2022)\citenamefont {Schlimgen}, \citenamefont {Head-Marsden}, \citenamefont {Sager}, \citenamefont {Narang},\ and\ \citenamefont {Mazziotti}}]{schlimgen_quantum_2022}%
  \BibitemOpen
  \bibfield  {author} {\bibinfo {author} {\bibfnamefont {A.~W.}\ \bibnamefont {Schlimgen}}, \bibinfo {author} {\bibfnamefont {K.}~\bibnamefont {Head-Marsden}}, \bibinfo {author} {\bibfnamefont {L.~M.}\ \bibnamefont {Sager}}, \bibinfo {author} {\bibfnamefont {P.}~\bibnamefont {Narang}},\ and\ \bibinfo {author} {\bibfnamefont {D.~A.}\ \bibnamefont {Mazziotti}},\ }\href {https://doi.org/10.1103/PhysRevResearch.4.023216} {\bibfield  {journal} {\bibinfo  {journal} {Physical Review Research}\ }\textbf {\bibinfo {volume} {4}},\ \bibinfo {pages} {023216} (\bibinfo {year} {2022})}\BibitemShut {NoStop}%
\bibitem [{\citenamefont {{Matthew Pocrnic}}\ \emph {et~al.}(2023)\citenamefont {{Matthew Pocrnic}}, \citenamefont {{D. Segal}},\ and\ \citenamefont {{Nathan Wiebe}}}]{matthew_pocrnic_quantum_2023}%
  \BibitemOpen
  \bibfield  {author} {\bibinfo {author} {\bibnamefont {{Matthew Pocrnic}}}, \bibinfo {author} {\bibnamefont {{D. Segal}}},\ and\ \bibinfo {author} {\bibnamefont {{Nathan Wiebe}}},\ }\href@noop {} {\bibfield  {journal} {\bibinfo  {journal} {Quantum Science and Technology}\ } (\bibinfo {year} {2023})}\BibitemShut {NoStop}%
\bibitem [{\citenamefont {Moss}\ \emph {et~al.}(2023)\citenamefont {Moss}, \citenamefont {Ebadi}, \citenamefont {Wang}, \citenamefont {Semeghini}, \citenamefont {Bohrdt}, \citenamefont {Lukin},\ and\ \citenamefont {Melko}}]{moss_enhancing_2023}%
  \BibitemOpen
  \bibfield  {author} {\bibinfo {author} {\bibfnamefont {M.~S.}\ \bibnamefont {Moss}}, \bibinfo {author} {\bibfnamefont {S.}~\bibnamefont {Ebadi}}, \bibinfo {author} {\bibfnamefont {T.~T.}\ \bibnamefont {Wang}}, \bibinfo {author} {\bibfnamefont {G.}~\bibnamefont {Semeghini}}, \bibinfo {author} {\bibfnamefont {A.}~\bibnamefont {Bohrdt}}, \bibinfo {author} {\bibfnamefont {M.~D.}\ \bibnamefont {Lukin}},\ and\ \bibinfo {author} {\bibfnamefont {R.~G.}\ \bibnamefont {Melko}},\ }\href {https://doi.org/10.48550/arXiv.2308.02647} {\bibinfo {title} {Enhancing variational {Monte} {Carlo} using a programmable quantum simulator}} (\bibinfo {year} {2023}),\ \bibinfo {note} {arXiv:2308.02647 [cond-mat, physics:quant-ph]}\BibitemShut {NoStop}%
\bibitem [{\citenamefont {Luo}\ \emph {et~al.}(2024)\citenamefont {Luo}, \citenamefont {Lin},\ and\ \citenamefont {Gao}}]{luo_variational_2024}%
  \BibitemOpen
  \bibfield  {author} {\bibinfo {author} {\bibfnamefont {J.}~\bibnamefont {Luo}}, \bibinfo {author} {\bibfnamefont {K.}~\bibnamefont {Lin}},\ and\ \bibinfo {author} {\bibfnamefont {X.}~\bibnamefont {Gao}},\ }\href {https://doi.org/10.1021/acs.jpclett.4c00576} {\bibfield  {journal} {\bibinfo  {journal} {The Journal of Physical Chemistry Letters}\ }\textbf {\bibinfo {volume} {15}},\ \bibinfo {pages} {3516} (\bibinfo {year} {2024})}\BibitemShut {NoStop}%
\bibitem [{\citenamefont {Ding}\ \emph {et~al.}(2024)\citenamefont {Ding}, \citenamefont {Li},\ and\ \citenamefont {Lin}}]{ding_simulating_2024}%
  \BibitemOpen
  \bibfield  {author} {\bibinfo {author} {\bibfnamefont {Z.}~\bibnamefont {Ding}}, \bibinfo {author} {\bibfnamefont {X.}~\bibnamefont {Li}},\ and\ \bibinfo {author} {\bibfnamefont {L.}~\bibnamefont {Lin}},\ }\href {https://doi.org/10.1103/PRXQuantum.5.020332} {\bibfield  {journal} {\bibinfo  {journal} {PRX Quantum}\ }\textbf {\bibinfo {volume} {5}},\ \bibinfo {pages} {020332} (\bibinfo {year} {2024})}\BibitemShut {NoStop}%
\bibitem [{\citenamefont {Olaya-Agudelo}\ \emph {et~al.}(2024)\citenamefont {Olaya-Agudelo}, \citenamefont {Stewart}, \citenamefont {Valahu}, \citenamefont {MacDonell}, \citenamefont {Millican}, \citenamefont {Matsos}, \citenamefont {Scuccimarra}, \citenamefont {Tan},\ and\ \citenamefont {Kassal}}]{olaya-agudelo_simulating_2024}%
  \BibitemOpen
  \bibfield  {author} {\bibinfo {author} {\bibfnamefont {V.~C.}\ \bibnamefont {Olaya-Agudelo}}, \bibinfo {author} {\bibfnamefont {B.}~\bibnamefont {Stewart}}, \bibinfo {author} {\bibfnamefont {C.~H.}\ \bibnamefont {Valahu}}, \bibinfo {author} {\bibfnamefont {R.~J.}\ \bibnamefont {MacDonell}}, \bibinfo {author} {\bibfnamefont {M.~J.}\ \bibnamefont {Millican}}, \bibinfo {author} {\bibfnamefont {V.~G.}\ \bibnamefont {Matsos}}, \bibinfo {author} {\bibfnamefont {F.}~\bibnamefont {Scuccimarra}}, \bibinfo {author} {\bibfnamefont {T.~R.}\ \bibnamefont {Tan}},\ and\ \bibinfo {author} {\bibfnamefont {I.}~\bibnamefont {Kassal}},\ }\href {https://doi.org/10.48550/arXiv.2407.17819} {\bibinfo {title} {Simulating open-system molecular dynamics on analog quantum computers}} (\bibinfo {year} {2024}),\ \bibinfo {note} {arXiv:2407.17819 [physics, physics:quant-ph]}\BibitemShut {NoStop}%
\bibitem [{\citenamefont {Trivedi}\ \emph {et~al.}(2024)\citenamefont {Trivedi}, \citenamefont {Franco~Rubio},\ and\ \citenamefont {Cirac}}]{trivedi_quantum_2024}%
  \BibitemOpen
  \bibfield  {author} {\bibinfo {author} {\bibfnamefont {R.}~\bibnamefont {Trivedi}}, \bibinfo {author} {\bibfnamefont {A.}~\bibnamefont {Franco~Rubio}},\ and\ \bibinfo {author} {\bibfnamefont {J.~I.}\ \bibnamefont {Cirac}},\ }\href {https://doi.org/10.1038/s41467-024-50750-x} {\bibfield  {journal} {\bibinfo  {journal} {Nature Communications}\ }\textbf {\bibinfo {volume} {15}},\ \bibinfo {pages} {6507} (\bibinfo {year} {2024})}\BibitemShut {NoStop}%
\bibitem [{\citenamefont {{David Layden}}\ \emph {et~al.}(2023)\citenamefont {{David Layden}}, \citenamefont {{Guglielmo Mazzola}}, \citenamefont {{Ryan V. Mishmash}}, \citenamefont {{Mário Motta}}, \citenamefont {{Paweł Wocjan}}, \citenamefont {{Jin-Sung Kim}},\ and\ \citenamefont {{Sarah Sheldon}}}]{david_layden_quantum-enhanced_2023}%
  \BibitemOpen
  \bibfield  {author} {\bibinfo {author} {\bibnamefont {{David Layden}}}, \bibinfo {author} {\bibnamefont {{Guglielmo Mazzola}}}, \bibinfo {author} {\bibnamefont {{Ryan V. Mishmash}}}, \bibinfo {author} {\bibnamefont {{Mário Motta}}}, \bibinfo {author} {\bibnamefont {{Paweł Wocjan}}}, \bibinfo {author} {\bibnamefont {{Jin-Sung Kim}}},\ and\ \bibinfo {author} {\bibnamefont {{Sarah Sheldon}}},\ }\href {https://doi.org/10.1038/s41586-023-06095-4} {\bibfield  {journal} {\bibinfo  {journal} {Nature}\ }\textbf {\bibinfo {volume} {619}},\ \bibinfo {pages} {282} (\bibinfo {year} {2023})}\BibitemShut {NoStop}%
\bibitem [{\citenamefont {Fioroni}\ \emph {et~al.}(2025)\citenamefont {Fioroni}, \citenamefont {Rojkov},\ and\ \citenamefont {Reiter}}]{fioroni_learning_2025}%
  \BibitemOpen
  \bibfield  {author} {\bibinfo {author} {\bibfnamefont {L.}~\bibnamefont {Fioroni}}, \bibinfo {author} {\bibfnamefont {I.}~\bibnamefont {Rojkov}},\ and\ \bibinfo {author} {\bibfnamefont {F.}~\bibnamefont {Reiter}},\ }\href {https://doi.org/10.48550/arXiv.2501.05350} {\bibinfo {title} {A learning agent-based approach to the characterization of open quantum systems}} (\bibinfo {year} {2025}),\ \bibinfo {note} {arXiv:2501.05350}\BibitemShut {NoStop}%
\bibitem [{\citenamefont {{Fabrizio Minganti}}\ \emph {et~al.}(2018)\citenamefont {{Fabrizio Minganti}}, \citenamefont {Minganti}, \citenamefont {{Alberto Biella}}, \citenamefont {Biella}, \citenamefont {{N. Bartolo}}, \citenamefont {Bartolo}, \citenamefont {{Cristiano Ciuti}},\ and\ \citenamefont {Ciuti}}]{fabrizio_minganti_spectral_2018}%
  \BibitemOpen
  \bibfield  {author} {\bibinfo {author} {\bibnamefont {{Fabrizio Minganti}}}, \bibinfo {author} {\bibfnamefont {F.}~\bibnamefont {Minganti}}, \bibinfo {author} {\bibnamefont {{Alberto Biella}}}, \bibinfo {author} {\bibfnamefont {A.}~\bibnamefont {Biella}}, \bibinfo {author} {\bibnamefont {{N. Bartolo}}}, \bibinfo {author} {\bibfnamefont {N.}~\bibnamefont {Bartolo}}, \bibinfo {author} {\bibnamefont {{Cristiano Ciuti}}},\ and\ \bibinfo {author} {\bibfnamefont {C.}~\bibnamefont {Ciuti}},\ }\href {https://doi.org/10.1103/physreva.98.042118} {\bibfield  {journal} {\bibinfo  {journal} {Physical Review A}\ }\textbf {\bibinfo {volume} {98}},\ \bibinfo {pages} {042118} (\bibinfo {year} {2018})}\BibitemShut {NoStop}%
\bibitem [{Note5()}]{Note5}%
  \BibitemOpen
  \bibinfo {note} {It may also have purely imaginary eigenvalues representing a limit cycle instead of a steady state, albeit not in the context we consider here.}\BibitemShut {Stop}%
\bibitem [{Note6()}]{Note6}%
  \BibitemOpen
  \bibinfo {note} {The $d=4$ generalized Gell Mann matrices are an alternative equivalent choice}\BibitemShut {NoStop}%
\bibitem [{\citenamefont {Pearson}(1901)}]{Pearson1901}%
  \BibitemOpen
  \bibfield  {author} {\bibinfo {author} {\bibfnamefont {K.}~\bibnamefont {Pearson}},\ }\href {https://doi.org/10.1080/14786440109462720} {\bibfield  {journal} {\bibinfo  {journal} {The London, Edinburgh, and Dublin Philosophical Magazine and Journal of Science}\ }\textbf {\bibinfo {volume} {2}},\ \bibinfo {pages} {559} (\bibinfo {year} {1901})},\ \Eprint {https://arxiv.org/abs/https://doi.org/10.1080/14786440109462720} {https://doi.org/10.1080/14786440109462720} \BibitemShut {NoStop}%
\bibitem [{\citenamefont {Lloyd}(1982)}]{Lloyd1982}%
  \BibitemOpen
  \bibfield  {author} {\bibinfo {author} {\bibfnamefont {S.}~\bibnamefont {Lloyd}},\ }\href {https://doi.org/10.1109/TIT.1982.1056489} {\bibfield  {journal} {\bibinfo  {journal} {IEEE Transactions on Information Theory}\ }\textbf {\bibinfo {volume} {28}},\ \bibinfo {pages} {129} (\bibinfo {year} {1982})}\BibitemShut {NoStop}%
\bibitem [{\citenamefont {Ketchen}\ and\ \citenamefont {Shook}(1996)}]{david_j_ketchen_application_1996}%
  \BibitemOpen
  \bibfield  {author} {\bibinfo {author} {\bibfnamefont {D.~J.}\ \bibnamefont {Ketchen}}\ and\ \bibinfo {author} {\bibfnamefont {C.~L.}\ \bibnamefont {Shook}},\ }\href {https://doi.org/10.1002/(sici)1097-0266(199606)17:6<441::aid-smj819>3.0.co;2-g} {\bibfield  {journal} {\bibinfo  {journal} {Strategic Management Journal}\ }\textbf {\bibinfo {volume} {17}},\ \bibinfo {pages} {441} (\bibinfo {year} {1996})}\BibitemShut {NoStop}%
\end{thebibliography}
\end{document}